\pdfoutput=1

\documentclass[11pt,twoside,a4paper,cmspaper,final,collab]{cms-tdr}
\def\svnVersion{f305a15-D}\def\svnDate{2020/09/25}\def\cmsCernNoTag{CERN-EP-2020-105}\def\cmsCernDate{\today}\def\cmsMessage{"Published in the Journal of Instrumentation as \href{http://dx.doi.org/10.1088/1748-0221/15/10/P10002}{\doi{10.1088/1748-0221/15/10/P10002}.}"}

\begin{document}\cmsNoteHeader{EGM-18-001}

\hyphenation{had-ron-i-za-tion}
\hyphenation{cal-or-i-me-ter}
\hyphenation{de-vices}

\cmsNoteHeader{EGM-18-001}
\title{Reconstruction of signal amplitudes in the CMS electromagnetic calorimeter in the presence of overlapping proton-proton interactions}

\date{\today}

\abstract{
A template fitting technique for reconstructing the amplitude of
signals produced by the lead tungstate crystals of the CMS
electromagnetic calorimeter is described. This novel approach is
designed to suppress the contribution to the signal of the increased
number of out-of-time interactions per beam crossing following the
reduction of the accelerator bunch spacing from 50 to 25\unit{ns} at
the start of Run~2 of the LHC.  Execution of the algorithm is
sufficiently fast for it to be employed in the CMS high-level trigger.
It is also used in the offline event reconstruction. Results obtained
from simulations and from Run~2 collision data (2015--2018)
demonstrate a substantial improvement in the energy resolution of the
calorimeter over a range of energies extending from a few GeV to
several tens of GeV.  }

\hypersetup{%
pdfauthor={CMS Collaboration},%
pdftitle={Reconstruction of signal amplitudes in the CMS electromagnetic calorimeter in the presence of overlapping proton-proton interactions},%
pdfsubject={CMS},%
pdfkeywords={CMS, ECAL, pattern recognition}}

\maketitle

\tableofcontents

\clearpage

\section{Introduction}
\label{sec:introduction}

The central feature of the CMS apparatus is a superconducting solenoid
of 6\unit{m} internal diameter, providing a magnetic field of
3.8\unit{T}. Within the solenoid volume are a silicon pixel and strip
tracker, a lead tungstate (PbWO$_4$) crystal electromagnetic
calorimeter (ECAL), which is the focus of this paper, and a brass and
scintillator hadron calorimeter (HCAL), each composed of a barrel and
two endcap sections. Forward calorimeters extend the pseudorapidity
coverage provided by the barrel and endcap detectors. Muons are
detected in gas-ionization chambers embedded in the steel flux-return
yoke outside the solenoid. A more detailed description of the CMS
detector is given in Ref.~\cite{Chatrchyan:2008zzk}.

The ECAL consists of 61\,200 PbWO$_4$ crystals mounted in the barrel
section (EB), covering the range of pseudorapidity $\abs{\eta}<1.48$,
closed by 7324 crystals in each of the two endcaps (EE), covering the
range $1.48 <\abs{\eta}< 3.0$.  The EB uses 23\unit{cm} long crystals
with front-face cross sections of approximately
$2.2{\times}2.2$\unit{cm$^2$}, while the EE contains 22\unit{cm} long
crystals with a front-face cross section of
$2.86{\times}2.86$\unit{cm$^2$}.  The scintillation light is detected
by avalanche photodiodes (APDs) in the EB and by vacuum phototriodes
(VPTs) in the EE.  The PbWO$_4$ crystals have a Moli\`ere radius of
2.19\unit{cm}, approximately matching the transverse dimensions of the
crystals.  A preshower detector consisting of two planes of silicon
sensors interleaved with lead for a total of 3 radiation lengths is
located in front of EE~\cite{CERN-LHCC-97-033}. A crystal transparency
monitoring system, based on the injection of laser light at
447\unit{nm}, close to the emission peak of scintillation light from
PbWO$_4$, is used to track and correct for response changes during LHC
operation~\cite{Anfreville:2007zz,Zhang:2005ip}.

The LHC operating conditions during Run~2 data taking (2015--2018)
were more challenging than those of Run~1 (2010--2013) in several
respects. The center-of-mass energy of the collisions was raised from
8 to 13\TeV, the bunch spacing (the time interval between neighboring
bunches), was halved from 50\unit{ns} to the design value of
25\unit{ns}, and the instantaneous luminosity reached
$2.1{\times}10^{34}\percms$ compared to $0.75{\times}10^{34}\percms$
achieved in 2012.

The mean number of additional interactions in a single bunch crossing
(BX), termed pileup (PU), in Run~2 was 34, with the tail of the
distribution extending up to 80. The average values for 2016, 2017 and
2018 were 27, 38 and 37, respectively. For the results shown in this
paper, obtained from simulations, an average number of 40 interactions
per bunch crossing is used.  For comparison, during Run~1 in 2012, the
mean value was 21 interactions per BX, with an extreme value of
40. After shaping by the electronics, the ECAL signals extend over
several hundred nanoseconds. Consequently, the decrease in the LHC
bunch spacing from 50 to 25\unit{ns} results in an increased number of
overlapping signals from neighboring BXs, referred to as out-of-time
(OOT) pileup.  These spurious signals effectively add to the
electronic noise and degrade the energy resolution of the
calorimeter. To reduce these effects, an innovative ECAL amplitude
reconstruction procedure, based on a template fitting technique, named
``multifit'', was introduced in 2015, before the start of Run~2.  The
new algorithm replaces the one used during Run~1 (``weights''
method)~\cite{Bruneliere:2006ra}, which was based on a
digital-filtering technique. The original algorithm performed well
under the conditions of Run~1, but was not suitable for Run~2 because
of the increased OOT pileup.

\section{Data and simulated samples}
\label{sec:dataSimulation}
The results shown in this paper are based on subsets of the data
samples recorded by the CMS experiment in proton-proton ($\Pp\Pp$)
collisions at a center-of-mass energy of 13\TeV. Calibration samples
are recorded by using special data streams, based either on a minimal
single-crystal energy deposit, or on diphoton invariant mass, to
profit from the copious production of $\PGpz$ mesons subsequently
decaying into $\gamma\gamma$.  Performances are evaluated on a subset
of the standard physics stream of the high-level trigger (HLT), by
using electrons from $\cPZ$-boson decays ($\cPZ\to\Pep\Pem$).

In addition to data samples from calibration sources and collision
data, two kinds of Monte Carlo (MC) samples are used. One is the full
detector simulation used for physics analyses, implemented
with \GEANTfour~\cite{geant4}, of single photons within the CMS
detector with a uniform distribution in $\eta$ and a flat transverse
momentum $\pt$ spectrum extending from 1 to 100\GeV.  These events are
generated with the $\PYTHIA$~8.226~\cite{Sjostrand:2014zea} package
and its CUETP8M1~\cite{Khachatryan:2015pea} tune for parton showering,
hadronization, and underlying event simulation. These events are used
to study the performance of the algorithm when the showering of an
electromagnetic particle spreads across more than a single crystal,
which is typical of most energy deposits in the ECAL.  The second set
of MC samples is produced by a fast stand-alone simulation, where the
single-crystal amplitudes are generated by pseudo-experiments using a
parametric representation of the pulse shape and the measured
covariance matrix. Energy deposits typical of the PU present in Run~2
are then added to these signals. Additional $\Pp\Pp$ interactions in
the same or adjacent BXs are added to each simulated event sample,
with an average number of 40.

\section{The electromagnetic calorimeter readout}
\label{sec:readout}

The electrical signal from the photodetectors is amplified and shaped
using a multigain preamplifier (MGPA), which provides three
simultaneous analogue outputs that are shaped to have a rise time of
approximately 50\unit{ns} and fall to 10\% of the peak value in
400\unit{ns}\cite{CERN-LHCC-97-033}. The shaped signals are sampled at
the LHC bunch-crossing frequency of 40\unit{MHz} and digitized by a
system of three channels of floating-point Analog-to-Digital
Converters (ADCs). The channel with the gain that gives the highest
nonsaturated value is selected sample-by-sample, thus providing a
dynamic range from 35\MeV to 1.7\TeV in the barrel. A time frame of 10
consecutive samples is read out every 25\unit{ns}, in synchronization
with the triggered LHC BX~\cite{CERN-LHCC-97-033}.  The convention
used throughout this report is to number samples starting from 0. The
phase of the readout is adjusted such that the time of the in-time
pulse maximum value coincides with the fifth digitized sample.  The
first three samples are read out before the signal pulse rises
significantly from the pedestal baseline (presamples).  The
50\unit{ns} rise time of the signal pulse after amplification results
from the convolution of the 10\unit{ns} decay time of the crystal
scintillation emission and the 40\unit{ns} shaping time of the
MGPA~\cite{CERN-LHCC-97-033,Bruneliere:2006ra,Chatrchyan:2008zzk}.

\section{The multifit method}
\label{sec:multifit}

\subsection{The Run~1 amplitude reconstruction of ECAL signals}
\label{sec:weights}

During LHC Run~1, a weighting algorithm~\cite{Bruneliere:2006ra} was
used to estimate the ECAL signal amplitudes, both online in the
HLT~\cite{Khachatryan:2016bia} and in the offline reconstruction.
With that algorithm the amplitude is estimated as a linear combination
of 10 samples, $s_i$:
\begin{linenomath} 
\begin{equation}
\label{eqn:weights}
\hat{\cal A}=\sum_{i=0}^{9} w_i s_i,
\end{equation}
\end{linenomath}
where the weights $w_i$ are calculated by minimizing the variance of
$\hat{\cal A}$. This algorithm was developed to provide an optimal
reduction of the electronics noise and a dynamic subtraction of the
pedestal, which is estimated on an event-by-event basis by the average
of the presamples.

The LHC Run~2 conditions placed stringent requirements on the ECAL
pulse reconstruction algorithm. Several methods were investigated to
mitigate the effect of the increased OOT pileup, to achieve optimal
noise performance. The methods that were studied included: using a
single sample at the signal pulse maximum, a deconvolution method
converting the discrete time signal into the frequency
domain~\cite{Gadomski:1992xu}, and the multifit. The first one uses a
minimal information from the pulse shape and, although being robust
against OOT pileup, results in a degradation of energy resolution for
most of the energy range below $\approx$100\GeV. The second was the
subject of a pilot study and was never fully developed. The last one
is the subject of this paper.

\subsection{The multifit algorithm}
\label{sec:method}

The multifit method uses a template fit with $N_\text{BX}$ parameters,
comprising one in-time (IT) and up to nine OOT amplitudes, up to five
occurring before, and up to four after the IT pulse:
$N_\text{BX}\in[1-10]$.  The fit minimizes the $\chi^2$ defined as:
\begin{linenomath} 
\begin{equation}
\label{eqn:chi2}
\chi^2 = \left(\sum_{j=0}^{N_\text{BX}} A_j\vec{p}_j - \vec{S}\right)^T
\mathbf{C}^{-1}
\left(\sum_{j=0}^{N_\text{BX}} A_j\vec{p}_j - \vec{S}\right),
\end{equation}
\end{linenomath} 
where the vector $\vec{S}$ comprises the 10 readout samples, $s_i$,
after having subtracted the pedestal value, $\vec{p}_j$ are the pulse
templates for each BX, and $A_{j}$, which are obtained by the fit, are
the signal pulse amplitudes in ten consecutive BXs, with $A_{5}$
corresponding to the IT BX.  The pulse templates $\vec{p}_j$ for each
BX have the same shape, but are shifted in time by $j$ multiples of
25\unit{ns}.  The pulse templates are described by binned
distributions with 15 bins of width 25\unit{ns}. An extension of five
additional time samples after the $10^{th}$ sample (the last digitized
one) is used to obtain an accurate description of the contribution to
the signal from early OOT pulses with tails that overlap the IT pulse.

The total covariance matrix $\mathbf{C}$ used in the $\chi^2$
minimization of Eq.~(\ref{eqn:chi2}) includes the correlation of the
noise and the signal between the different time samples.  It is
defined as the quadratic sum of two contributions:
\begin{linenomath} 
\begin{equation}
\label{eqn:totalcovariance}
\mathbf{C} = \mathbf{C}_\text{noise} \oplus \sum_{j=0}^{N_\text{BX}} A_j^2\mathbf{C}_{\text{pulse}}^j,
\end{equation}
\end{linenomath} 
where $\mathbf{C}_\text{noise}$ is the covariance matrix associated
with the electronics noise and $\mathbf{C}_{\text{pulse}}^j$ is the
one associated with the pulse shape template. Each channel of the
ECAL, \ie, a crystal with its individual readout, is assigned its own
covariance matrix.  Quadratic summation of the two components is
justified since the variance for the pulse templates is uncorrelated
with the electronic noise. In fact, the uncertainty in the shape of
the signal pulses for a given channel is dominated by event-by-event
fluctuations of the beam spot position along the $z$-axis, of order
several cms~\cite{TRK_001}, which affect the arrival time of particles
at the front face of ECAL.

The $\mathbf{C}_{\text{pulse}}$ matrix is
calculated as:
\begin{linenomath}
\begin{equation}
\label{eqn:covariance}
C_\text{pulse}^{i,k} = \frac{\sum_{n=1}^{N_\text{events}} \tilde{s}_i(n)\tilde{s}_k(n)} {N_\text{events}},
\end{equation}
\end{linenomath} 
where the $\tilde{s}_i(n)$ are the pedestal-subtracted sample values,
$s_i(n)-P$, scaled for each event $n$, such that $\tilde{s}_5(n) =
1$. The value of $P$ equals the average of the three unscaled
presamples over many events. Both the templates and their covariance
matrices are estimated from collision data and may vary with time, for
reasons described in Section~\ref{sec:templates}.  The electronics
noise dominates the uncertainty for low-energy pulses, whereas the
uncertainty in the template shape dominates for higher energies. The
determination of $\mathbf{C}_\text{noise}$, which is calculated
analogously as $\mathbf{C}_{\text{pulse}}$, but with dedicated data,
is described in Section~\ref{sec:noise}.

\begin{figure}[!t]
\centering
\includegraphics[width=0.49\textwidth]{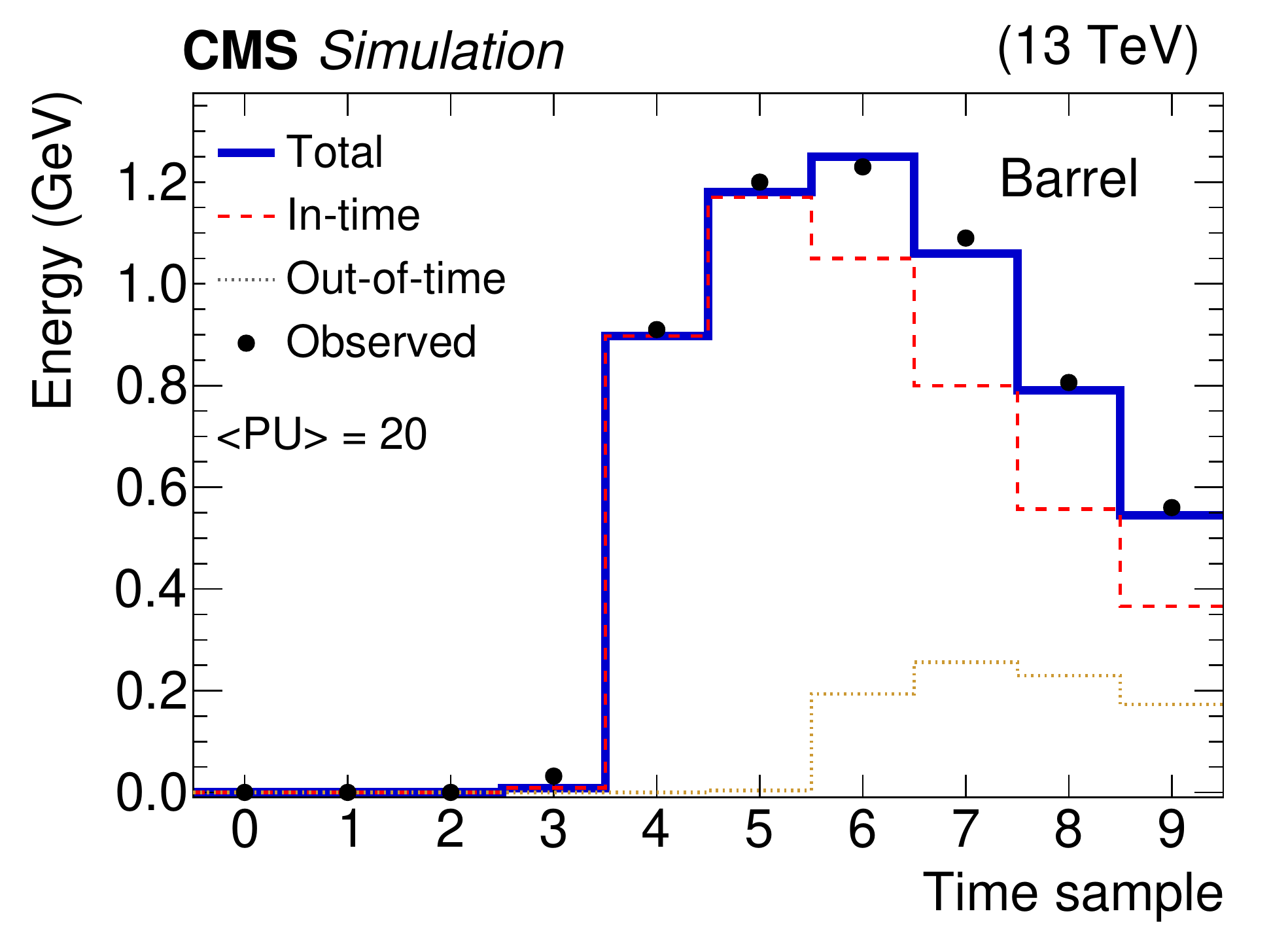}
\includegraphics[width=0.49\textwidth]{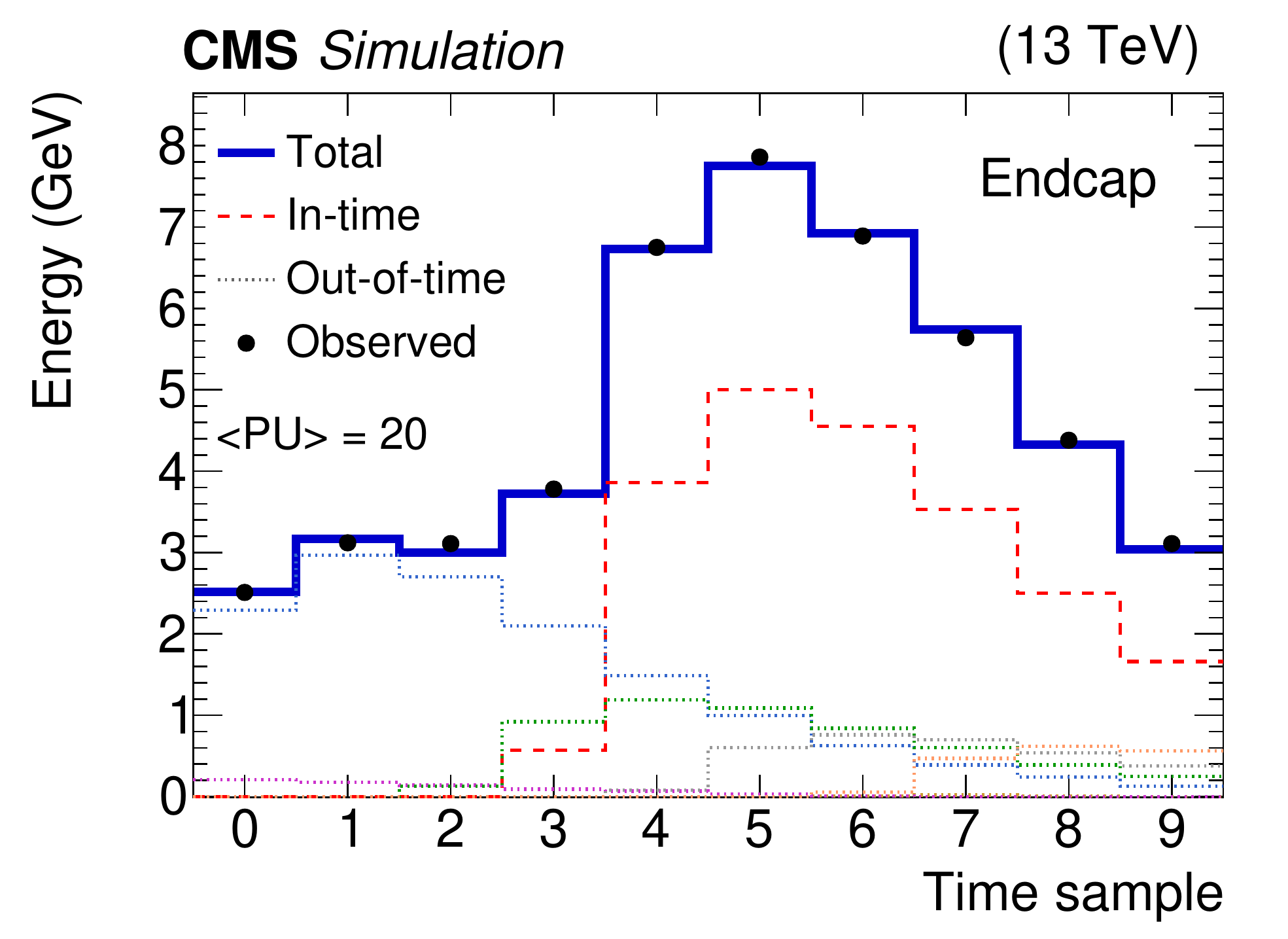}
\caption{Two  examples of fitted pulses for simulated events with 20 average
  pileup interactions and 25\unit{ns} bunch spacing. Signals from
  individual crystals are shown. They arise from a $\pt = 10\GeV$
  photon shower in the barrel (left) and in an endcap (right). In the
  left panel, one OOT pulse, in addition to the IT pulse, is fitted. In
  the right panel, six OOT pulses, in addition to the IT pulse, are
  fitted. Filled
  circles with error bars represent the 10 digitized samples, the red
  dashed distributions (dotted multicolored distributions) represent
  the fitted in-time (out-of time) pulses with positive
  amplitudes. The solid dark blue histograms represent the sum of all
  the fitted contributions. Within the dotted distributions, the color
  distinguishes the fitted out-of-time pulses with different BX, while
  the legend represent them as a generic gray dotted
  line.  \label{fig:multifits}}
\end{figure}

The minimization of the $\chi^2$ in Eq.~(\ref{eqn:chi2}) has to be
robust and fast to use both in the offline CMS reconstruction and at
the HLT. In particular, the latter has tight computation time
constraints, especially in the EB, where the number of channels that
are read out (typically 1000 and as high as 4000) for every triggered
BX, poses a severe limitation on the time allowed for each
minimization. Therefore, the possibility of using minimization
algorithms like \textsc{minuit} \cite{James:1975dr} to perform the
$10{\times}10$ matrix inversion is excluded.  Instead, the technique
of nonnegative least squares \cite{nnls} is used, with the constraint
that the fitted amplitudes $A_j$ are all positive.  The $\chi^2$
minimization is performed iteratively. First, all the amplitudes are
set to zero, and one nonzero amplitude at a time is added. The
evaluation of the inverse matrix $\mathbf{C}^{-1}$, which is the
computationally intensive operation, is iterated until the $\chi^2$
value in Eq.~(\ref{eqn:chi2}) converges
($\Delta\chi^2<10^{-3}$)~\cite{lawson}. Usually the convergence is
reached with fewer than 10 nonzero fitted amplitudes, so the system
is, in general, over-constrained. Examples of fitted pulses in single
crystals of the EB and EE are shown in Fig.~\ref{fig:multifits}
(right) and (left), respectively. They are obtained from a full
detector simulation of photons with transverse momentum $\pt=10\GeV$.

Since the only unknown quantities are the fitted amplitudes, the
minimization corresponds to the solution of a system of linear
equations with respect to a maximum of 10 nonnegative $A_j$
values. The implementation uses a C++ template linear algebra
library, \textsc{eigen}~\cite{eigenweb}, which is versatile and highly
optimized. The time required to compute the amplitude of all the
channels in one event is approximately 10\unit{ms} for typical Run~2
events where the bunch spacing was 25\unit{ns} and there is an average
of 40 PU interactions per BX. The timing has been measured on an Intel
Xeon E5-2650v2 processor, which was used for the benchmark tests of
the CMS HLT farm at the beginning of Run~2 in
2015~\cite{Richardson:2134639}. The CPU time needed is about 100 times
less than that which was used to perform the equivalent minimization
using \textsc{minuit}, and for all events is much less than the
maximum time of 100\unit{ms}/event allowed for the HLT.  The algorithm
implementation has also been adapted for execution on GPUs for the new
processor farm, which will be used for LHC Run~3, which is planned to
begin in 2022.

\section{Determination of the multifit parameters}
\label{sec:inputs}

\subsection{Pulse shape templates}
\label{sec:templates}

The templates for the $\vec{p}_j$ term in Eq.~(\ref{eqn:chi2}) are
histograms with 15 bins, and represent the expected time distribution
of a signal from an energy deposit in the ECAL crystals. The first 10
bins correspond to the samples that are read out in a triggered
event. Bins 10--14 describe the tail of the signal pulse shape.

The pulse template differs from crystal to crystal, both because of
intrinsic pulse shape differences and, more importantly, because of
differences in the relative time position of the pulse maximum,
$T_\text{max}$, between channels. The pulse templates have also been
found to vary with time, during Run~2, as a result of crystal
irradiation. Both of these effects must be corrected for, and the time
variation requires regular updates to the pulse shape templates during
data taking.

The pulse templates are constructed in situ from collision data
acquired with a zero-bias trigger, \ie, a beam activity
trigger~\cite{Khachatryan:2016bia}, and events recorded with a
dedicated high-rate calibration data stream~\cite{Chatrchyan:2013dga}.
In the calibration data stream, the ten digitized samples from all
single-crystal energy deposits above a predefined noise threshold are
recorded, while the rest of the event is dropped to limit the trigger
bandwidth. The energy deposits in these events receive contributions
from both IT and OOT interactions.  In a fraction of the LHC fills,
the circulating beams are configured so that a few of the bunch
collisions are isolated, \ie, occur between bunches that are not
surrounded by other bunches.  In these collisions, the nominal
single-bunch intensity is achieved without OOT pileup, so a special
trigger requirement to record them was developed.  This allows a clean
measurement of the templates of IT pulses only. An amplitude-weighted
average pulse template is obtained, and only hits with amplitudes
larger than approximately five times the root-mean-square spread of
the noise are used.

During 2017, the pulse templates were recalibrated about 30 times. The
LHC implemented collisions with isolated bunches only when the LHC was
not completely filled with bunches, during the intensity ramp up,
typically at the beginning of the yearly data taking and after each
technical stop, \ie, a scheduled period of several days without
collisions exploited by the LHC for accelerator developments. For all
other updates, normal bunch collisions were used.  For these, a
minimum amplitude threshold was imposed at the level of 1\GeV, or
$5\sigma_\text{noise}$ when this was greater, and the
amplitude-weighted average of the templates suppressed the relative
contribution of OOT PU pulses.  It was verified that the pulse
templates derived from isolated bunches are consistent with those
obtained from nonisolated bunches.  Anomalous signals in the APDs,
which have a distorted pulse shape, are rejected on the basis of the
single-crystal timing and the spatial distribution of the energy
deposit among neighboring
crystals~\cite{Petyt:2012sf,Chatrchyan:2013dga}.

\begin{figure}[!htbp]
\centering
\includegraphics[width=0.49\textwidth]{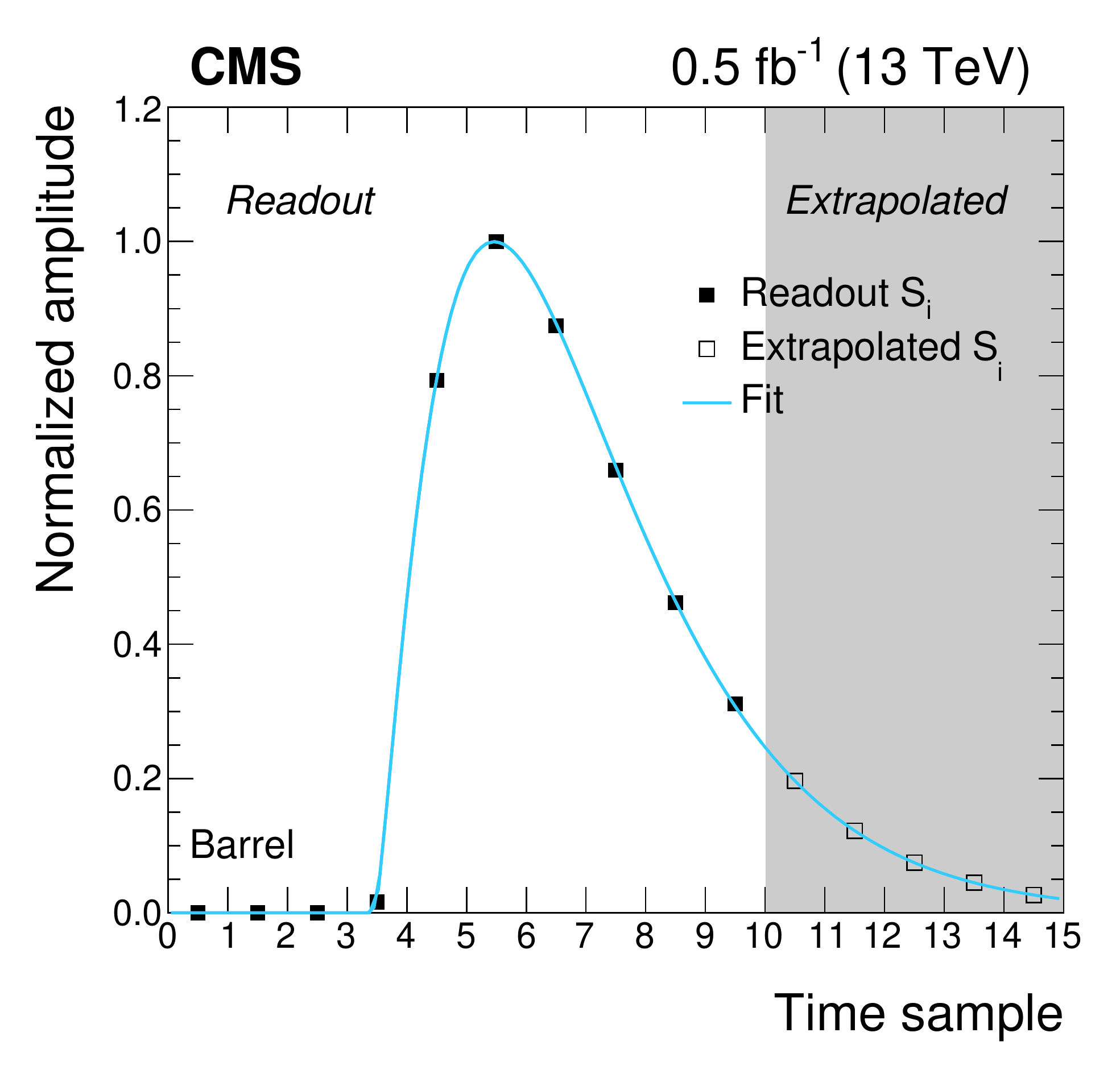}
\includegraphics[width=0.49\textwidth]{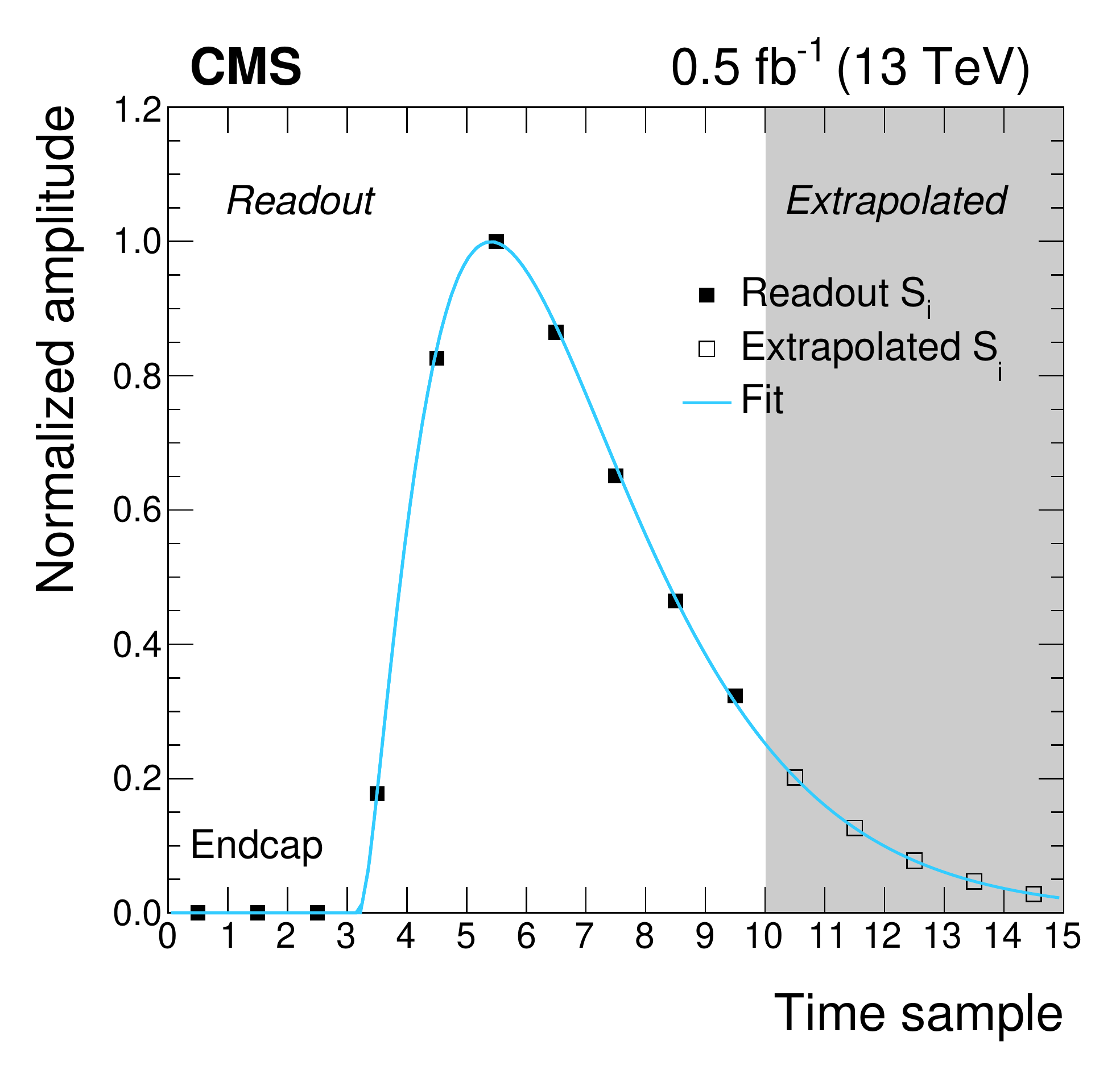}
  \caption{Pulse shape binned templates, measured in collision data
  recorded during June 2017 in a typical LHC fill, for a channel in
  the barrel (left) and in an endcap (right). The first 3 bins are the pedestal
  samples, and their values equal zero by construction.  The following
  7 bins are estimated from the average of the digitized samples on
  many hits, while the rightmost 5 bins are estimated by extrapolating
  the distribution using the function of Eq.~(\ref{eqn:alphabeta})
  (blue solid line). \label{fig:pulsetemplates}}
\end{figure}

The average pulse shape measured in the digitized time window of 10
BXs is extended by five additional time samples to model the falling
tail of the pulse, which is used to fit for the contribution of early
OOT pileup. This is achieved by fitting the average template with a
function of the form~\cite{Chatrchyan:2009qm}:
\begin{linenomath} 
\begin{equation}
\label{eqn:alphabeta}
A(t) = A\left(1+\frac{\Delta t}{\alpha\beta}\right)^\alpha \re^{-\Delta t/\beta}
\end{equation}
\end{linenomath} 
where $A$ represents the hit amplitude, $\Delta t=t-T_\text{max}$ the
time position relative to the peak, and $\alpha$, $\beta$ are two
shape parameters.  Examples of two average pulse shapes, obtained
using this method, are shown in Fig.~\ref{fig:pulsetemplates}.  The
extrapolation of the pulse templates outside of the readout window was
checked by injecting laser light into the crystals, with a shifted
readout phase. The tail of the pulse, measured in this way, agrees
with the extrapolated templates.

\begin{figure}[!htbp]
\centering
\includegraphics[width=0.49\textwidth]{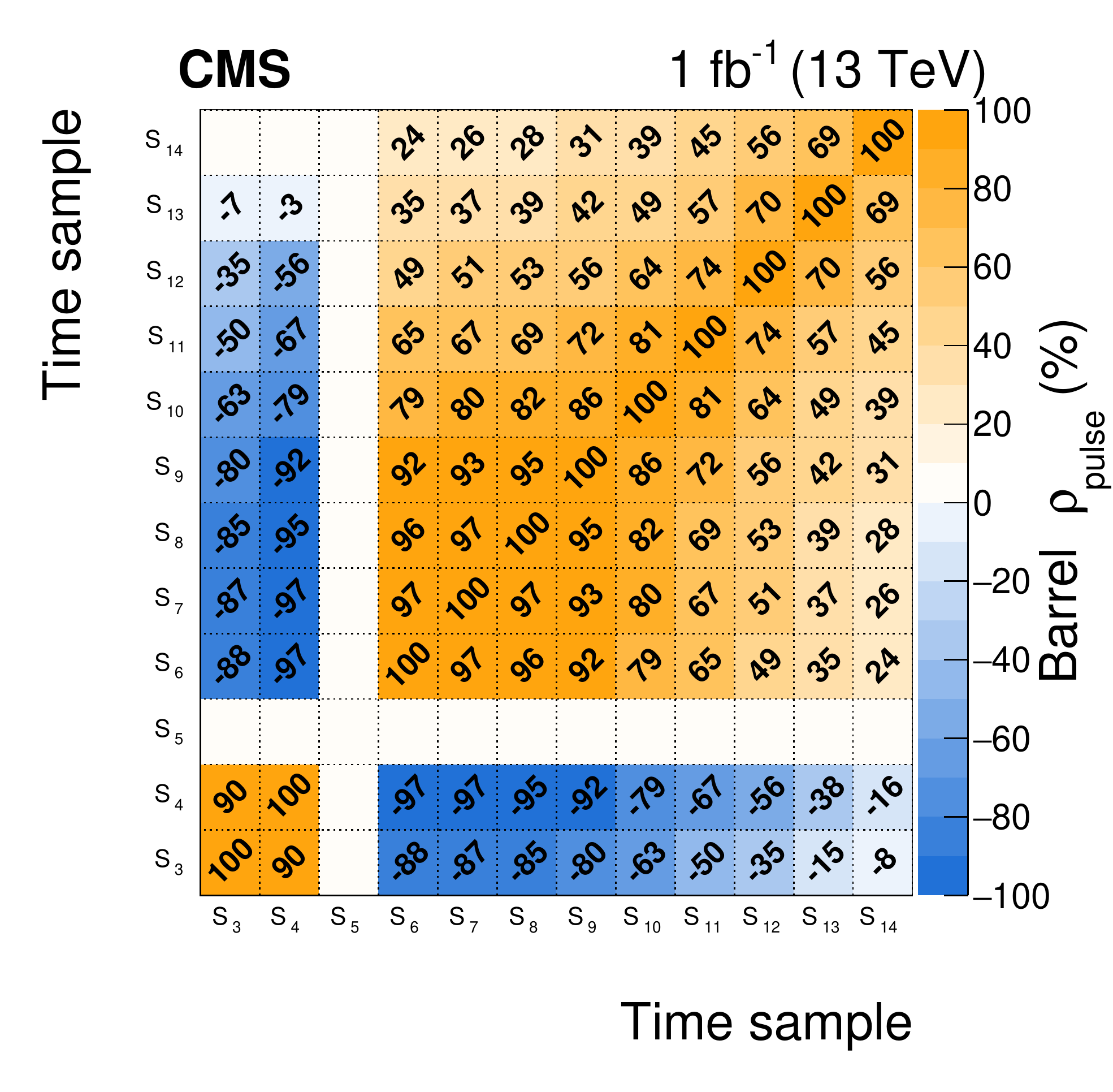}
\includegraphics[width=0.49\textwidth]{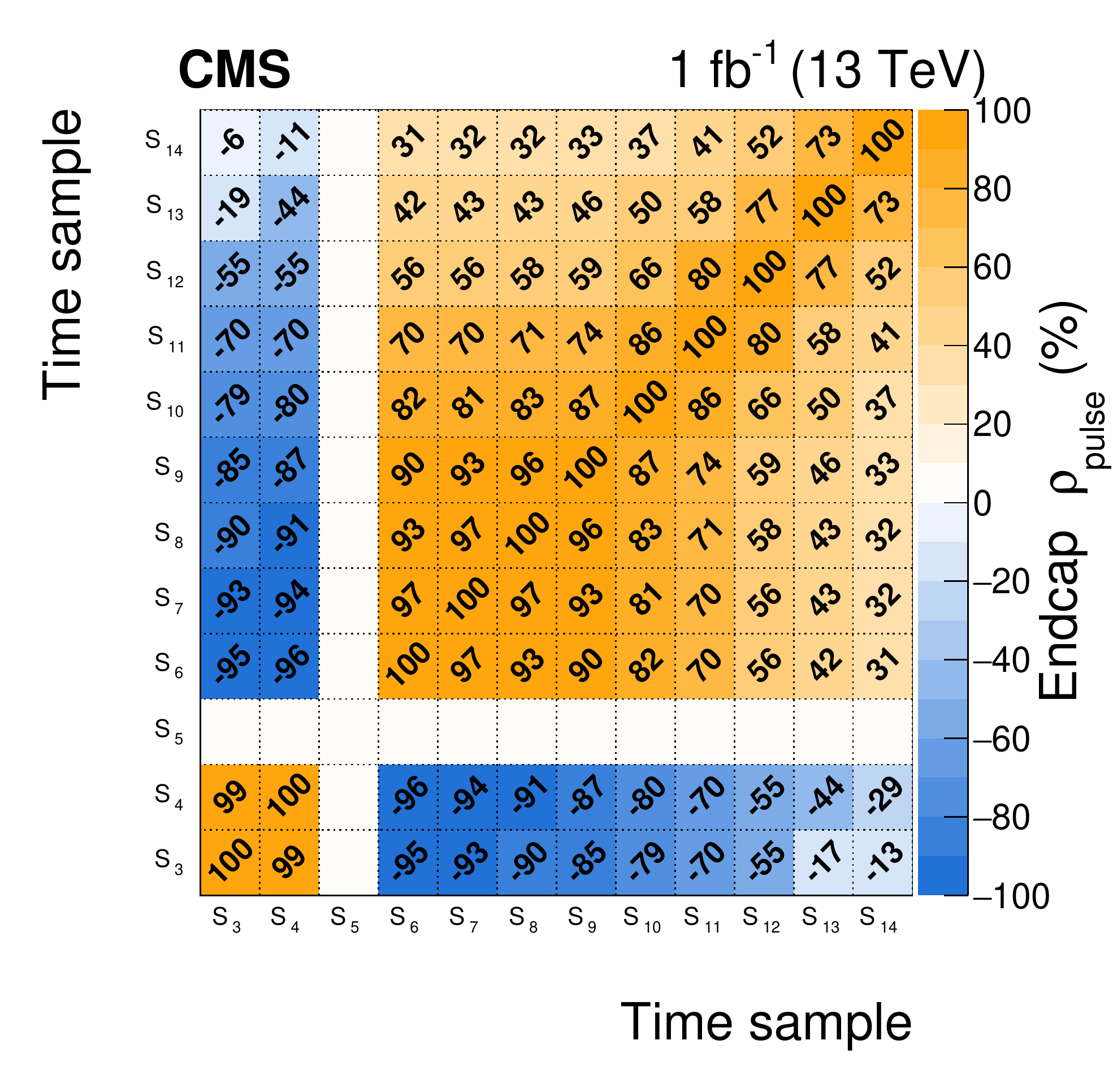}
  \caption{Correlation matrix of the pulse shape binned templates,
  $\boldsymbol{\rho}_\text{pulse}$, measured in collision data
  recorded during June 2017 in a typical LHC fill, for one channel in
  the barrel (left) and in an endcap (right). The elements with $i=5$
  or $k=5$ have zero variance by definition, since $S_5 = 1$ for all
  the hits. The elements $\rho_\text{pulse}^{i,k}$ with $i<3$ or $k<3$
  are the presamples, where no signal is expected, and are set to
  zero. Those with $i > 9$ or $k > 9$ are estimated from simulations
  with a shifted BX. The others are measured in collision data, as
  described in the text. All the $\rho_\text{pulse}^{i,k}$ values in
  the figure are expressed in percent for
  legibility. \label{fig:pulsecovariance}}

\end{figure}

The covariance matrix associated with the pulse template,
$\mathbf{C}_{\text{pulse}}$, is computed using
Eq.~(\ref{eqn:covariance}), with the same sample of digitized
templates used to determine the average pulse template and with the
same normalization and weighting strategy.  The correlation matrix of
the pulse template, $\boldsymbol{\rho}_\text{pulse}$, shown in
Fig.~\ref{fig:pulsecovariance}, is defined as
$\rho^{i,k}_\text{pulse}=C^{i,k}_\text{pulse}/(\sigma^i_\text{pulse}\sigma^k_\text{pulse})$,
where $\sigma^{i,k}_\text{pulse}$ is the square root of the variance
of the pulse shape for the $i,k$ bin of the template. The values of
$\sigma^i_\text{pulse}$ are in the range
$\left[5{\times}10^{-4}-1{\times}10^{-3}\right]$, the largest values
relative to samples in the tail of the pulse template.  The elements
of the covariance matrix outside the digitization window,
${C_\text{pulse}}^{i,k}$ with $i>9$ or $k>9$, are estimated from
simulations of single-photon events with the interaction time shifted
by an integer number of BXs. It was checked that this simulation
reproduces well the covariance matrix for the samples inside the
readout window.

The $\mathbf{C}_\text{pulse}$ matrix shows a strong correlation
between the time samples within either the rising edge or the falling
tail of the pulse. An anti-correlation is also observed between the
time samples of the rising edge and of the falling tail that is mostly
due to the spread in the particle arrival time at the ECAL surface,
which reflects the spatial and temporal distribution of the LHC beam
spot in CMS~\cite{CMS:2013gfa}. For the measured samples, the
correlations between $S_9$ and $S_8,S_7,S_6$, are all close to 1, with
values in the range (0.90--0.97). For the extrapolated samples, the
correlations change from bin to bin: between $S_{14}$ and
$S_{13},S_{12},S_{11}$ they are 0.69, 0.56, 0.45, respectively, in the
case of the barrel.

\subsection{Pedestals and electronic noise}
\label{sec:noise}

The pedestal mean is used in the multifit method to compute the
pedestal-subtracted template amplitudes $A_j$ in
Eq.~(\ref{eqn:chi2}). A bias in its measurement would reflect
almost linearly in a bias of the fitted amplitude, as discussed in
Section~\ref{sec:sensitivity}.

The covariance matrix associated with the electronic noise enters the
total covariance matrix of Eq.~(\ref{eqn:totalcovariance}).  It is
constructed as
$\mathbf{C}_\text{noise}=\sigma^2_\text{noise}\boldsymbol{\rho}_\text{noise}$,
where $\sigma_\text{noise}$ is the measured single-sample noise of the
channels, and $\boldsymbol{\rho}_\text{noise}$ is the noise
correlation matrix. The $\mathbf{C}_\text{noise}$ is calculated with
Eq.~(\ref{eqn:covariance}), where $i,k$ are the sample indices,
$\tilde{s}_{i}$ and $\tilde{s}_{j}$ are the measured sample values,
normalized to $\tilde{s}_5$, and $P$ is the expected value in the
absence of any signal, calculated, as for Eq.~(\ref{eqn:covariance}),
by averaging the three unscaled presamples over many events. Each
element of the noise covariance matrix is the mean over a large number
of events. The noise correlation matrix is defined as:
\begin{linenomath} 
\begin{equation}
\label{eqn:corrnoise}
\rho_\text{noise}^{i,k} = \frac{C^{i,k}_\text{noise}}{\sigma_\text{noise}^i\sigma_\text{noise}^k}.
\end{equation}
\end{linenomath} 
The average pedestal value and the electronic noise are measured
separately for the three MGPA gains.  For the highest gain value, data
from empty LHC bunches~\cite{Anfreville:2007zz,Zhang:2005ip} are used.
These are obtained by injecting laser light into the ECAL crystals in
coincidence with the bunch crossings. This gain value is used for the
vast majority of the reconstructed pulses (up to 150\GeV), and is very
sensitive to the electronics noise. One measurement per channel is
acquired approximately every 40 minutes. For the other two MGPA gains,
the pedestal mean and its fluctuations are measured from dedicated
runs without LHC beams present.

\begin{figure}[!htbp]
\centering
\includegraphics[width=0.49\textwidth]{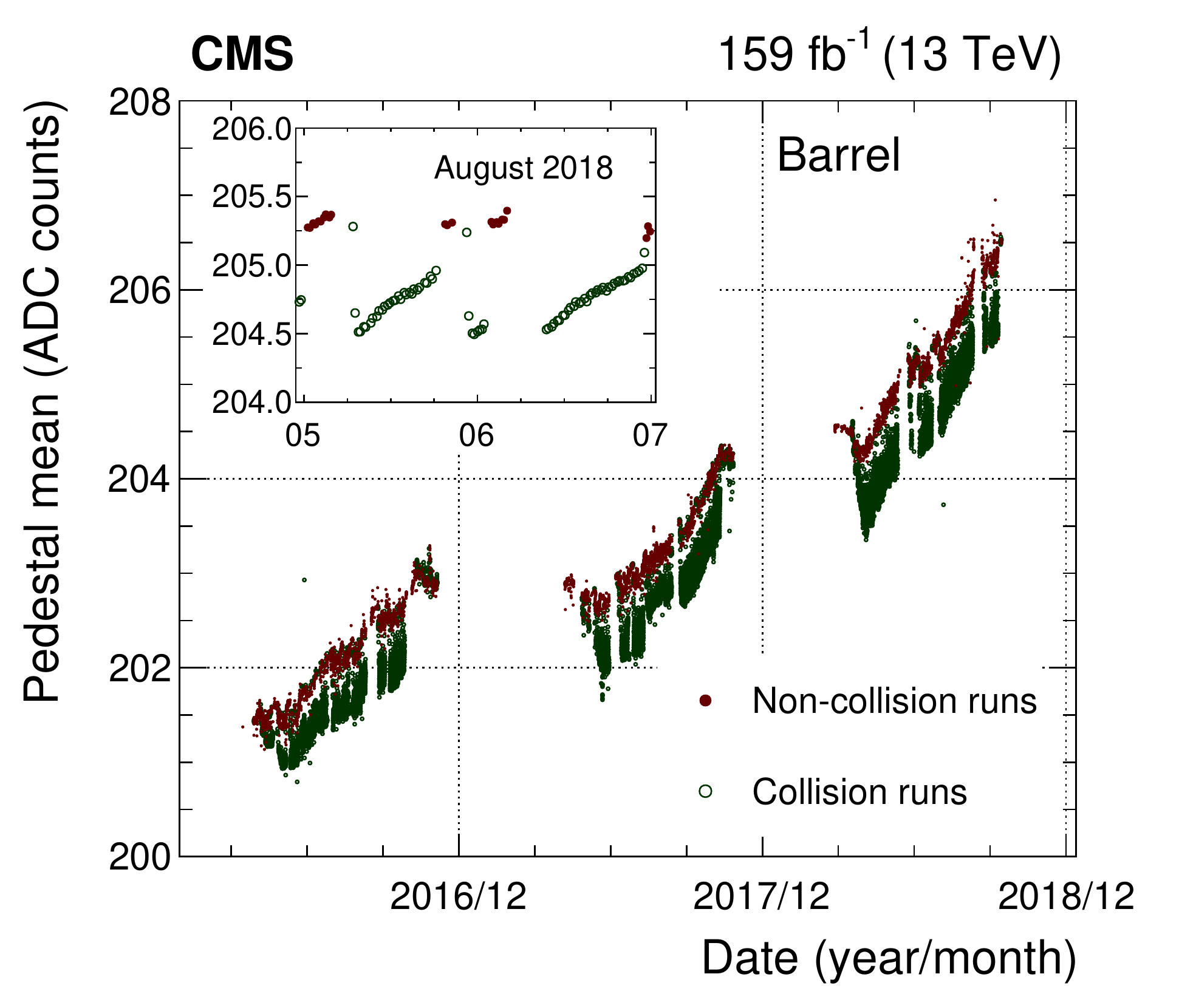}
\includegraphics[width=0.49\textwidth]{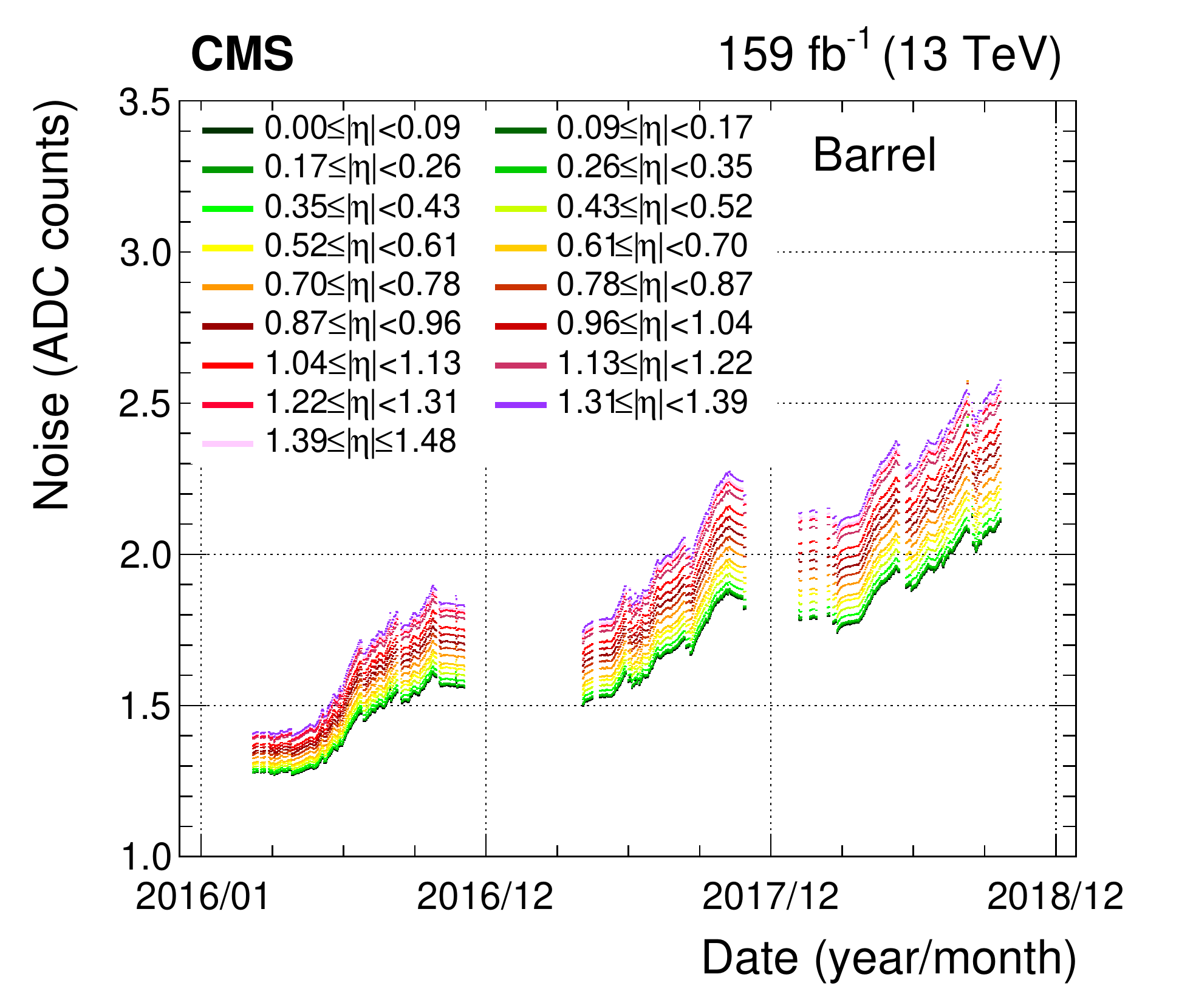}
  \caption{History of the pedestal mean value for the ECAL barrel
  (left) and its noise (right), measured for the highest MGPA gain in
  collision or noncollision runs taken during the 2016--2018 data
  taking period. The inset in the left panel shows an enlargement of
  two days in August 2018, to show in more detail the variation of the
  pedestal mean during LHC fills. \label{fig:pedhisto}}

\end{figure}

The time evolution of the pedestal mean in the EB during Run~2 is
shown for the highest MGPA gain in Fig.~\ref{fig:pedhisto} (left).  A
long-term, monotonic drift upwards is visible.  Short term (interfill)
luminosity related effects are also visible. The short-term variations
are smaller when the LHC luminosity is lower.  The long-term drift
depends on the integrated luminosity, while the short-term effects
depend on the instantaneous luminosity, and related to variations
inside the readout electronics. The behavior of the variation of the
pedestal value with time is similar at any $\abs{\eta}$ of the
crystal, while the magnitude of it increases with the
pseudorapidity, reflecting the higher irradiation.

The evolution of the electronic noise in the barrel is shown in
Fig.~\ref{fig:pedhisto} (right). It shows a monotonic increase with
time, related to the increase of the APD dark current due to the
larger radiation dose; no short-term luminosity-related effects are
visible.  For the barrel, where 1\unit{ADC} count $\cong$ 40\MeV, this
translates to an energy-equivalent noise of about 65\MeV at the
beginning of 2017 and 80\MeV at the end of the proton-proton running
in the same year.  A small decrease in the noise induced by the APD
dark current is visible after long periods without irradiation, \ie,
after the year-end LHC stops.  For the endcaps, the single-channel
noise related to the VPT signal does not evolve with time, and is
approximately 2\unit{ADC} counts. Nevertheless, the energy-equivalent
noise increases with time and with absolute pseudorapidity
$\abs{\eta}$ of the crystal because of the strong dependence of the
crystal transparency loss on $\abs{\eta}$ and time, due to higher
irradiation level. Consequently, the average noise at the end of 2017
in the endcaps translates to roughly 150\MeV up to
$\abs{\eta}\approx2$, whereas it increases to as much as 500\MeV at
the limit of the CMS tracker acceptance
($\abs{\eta}\approx2.5$). Thus, the relative contribution of
$\mathbf{C}_\text{noise}$ in the total covariance matrix strongly
depends on $\abs{\eta}$. For hits with amplitude larger than
$\approx$20\unit{ADC} counts, equivalent to an energy $\approx$1\GeV
before applying the light transparency corrections,
$\mathbf{C}_\text{pulse}$ dominates the covariance matrix for the
whole ECAL.

The covariance matrix for the noise used in the multifit is obtained
by multiplying the time independent correlation matrix in
Eq.~(\ref{eqn:corrnoise}) by the time dependent squared single sample
noise, $\sigma^2_\text{noise}$. The time evolution is automatically
accounted for by updating the values in the conditions
database~\cite{Guida:2015gvw}, with the measurements obtained in situ.

\begin{figure}[!htbp]
\centering
\includegraphics[width=0.49\textwidth]{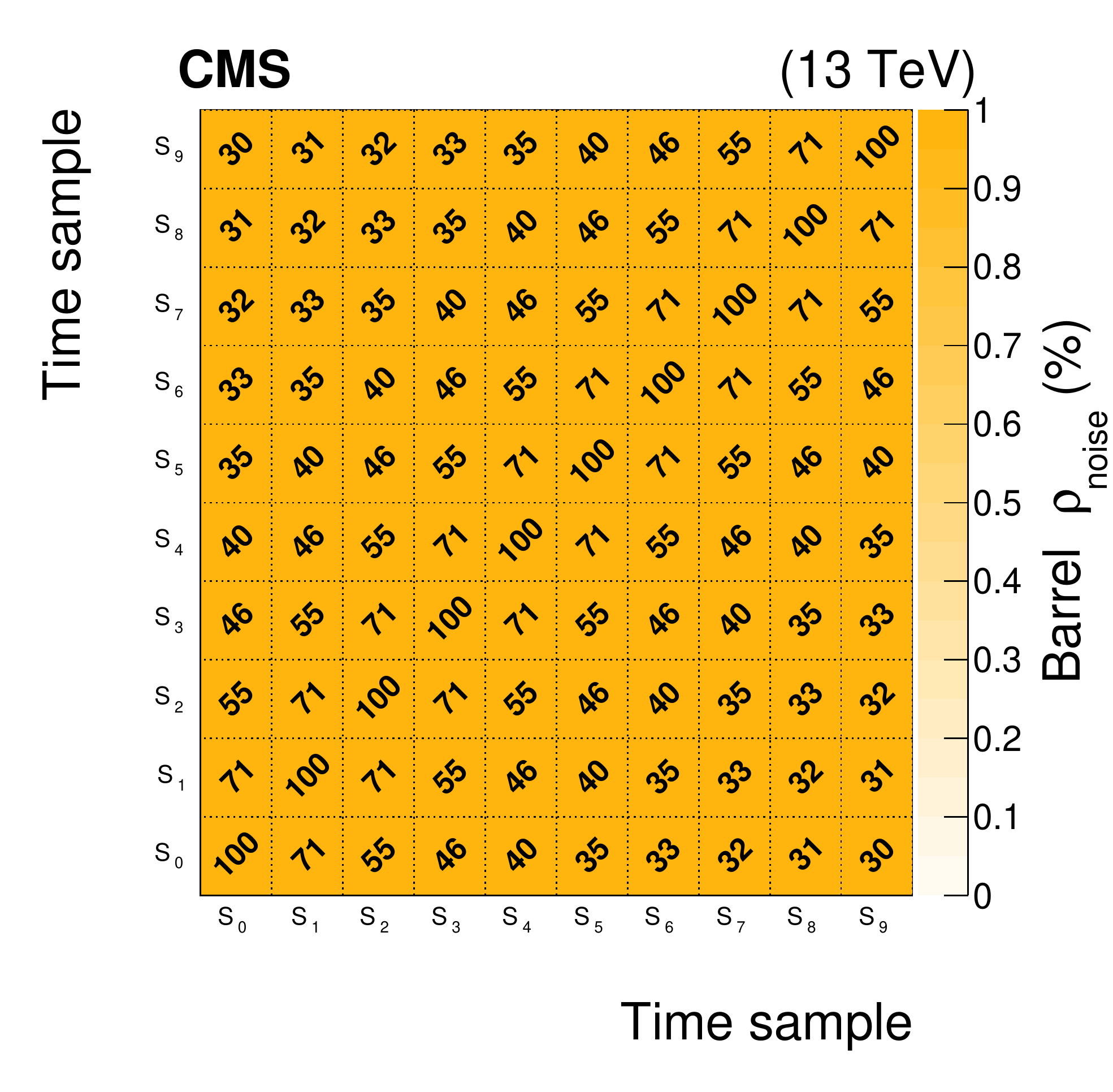}
\includegraphics[width=0.49\textwidth]{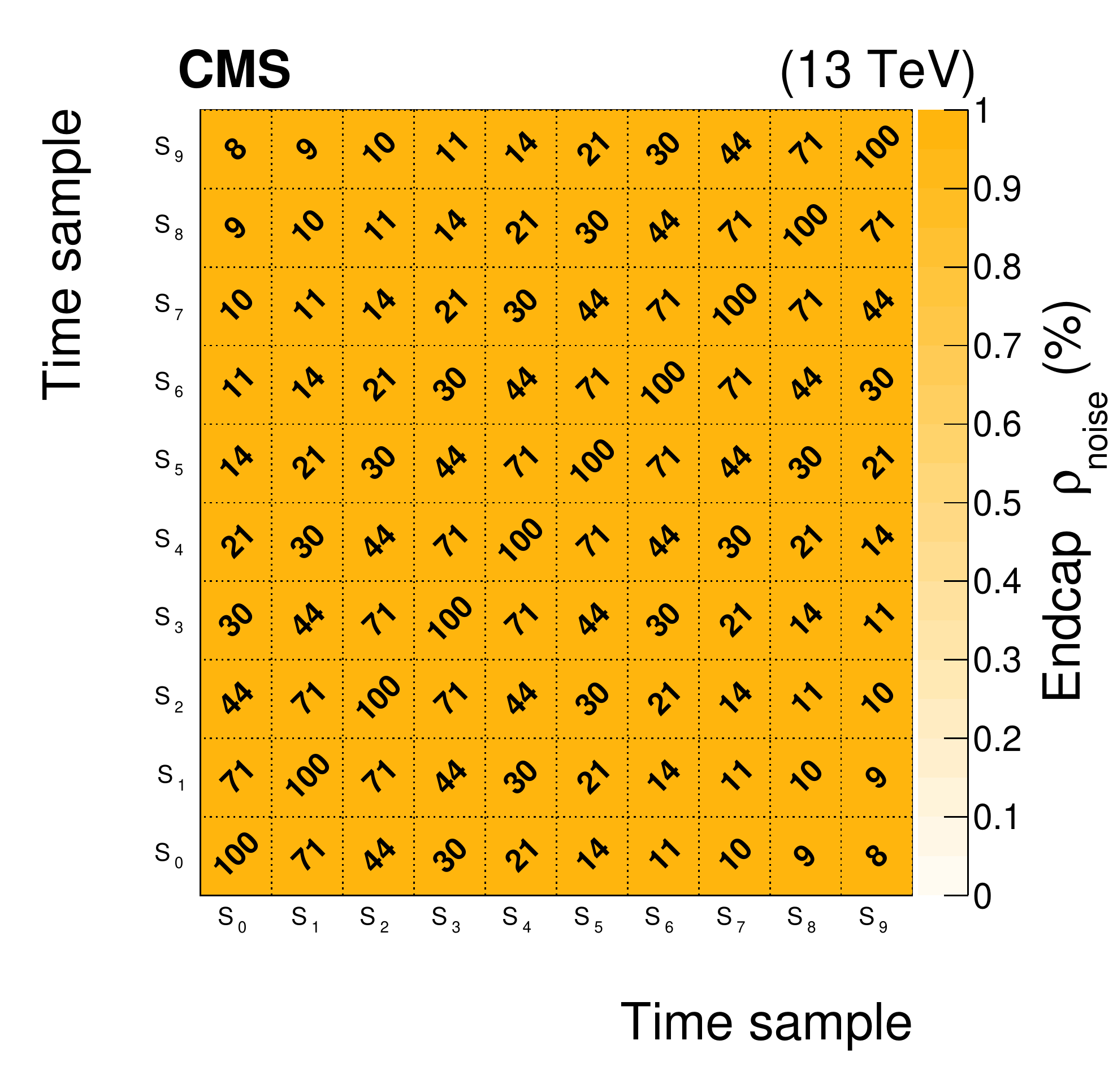}
  \caption{Correlation matrix of the electronics noise,
  $\boldsymbol{\rho}_\text{noise}$, measured in dedicated pedestal
  runs in Run~2, averaged over all the channels of the barrel (left)
  or endcaps (right). All the $\rho_\text{noise}^{i,k}$ values in the
  figure are expressed in percent for
  legibility. \label{fig:noisecorrmat}}

\end{figure}

Correlations between samples exist because of (1) the presence of
low-frequency (less than 4\unit{MHz}) noise that has been observed
during CMS operation~\cite{Chatrchyan:2009qm}, and (2) the effect of
the feedback resistor in the MGPA circuit~\cite{Antunovic}. The
correlation matrix of the electronic noise was measured with dedicated
pedestal runs; it is very similar for all channels within either the
EB or the EE, and stable with time. Consequently, it has been averaged
over all the channels within a single subsystem. The matrix for the
highest gain of the MGPA is shown in Fig.~\ref{fig:noisecorrmat}.
The MGPA component of the noise is such that the correlation depends
almost solely on the time distance between the two samples, following
an exponential relationship. For $\Delta t>100$\unit{ns}, it flattens
to a plateau corresponding to the low frequency noise.

\section{Sensitivity of the amplitude reconstruction to \\pulse timing and pedestal drifts}
\label{sec:sensitivity}

The multifit amplitude reconstruction utilizes as inputs pedestal
baseline values and signal pulse templates that are determined from
dedicated periodic measurements. Thus, it is sensitive to their
possible changes with time.

\begin{figure}[!htbp]
  \centering \includegraphics[width=0.6\textwidth]{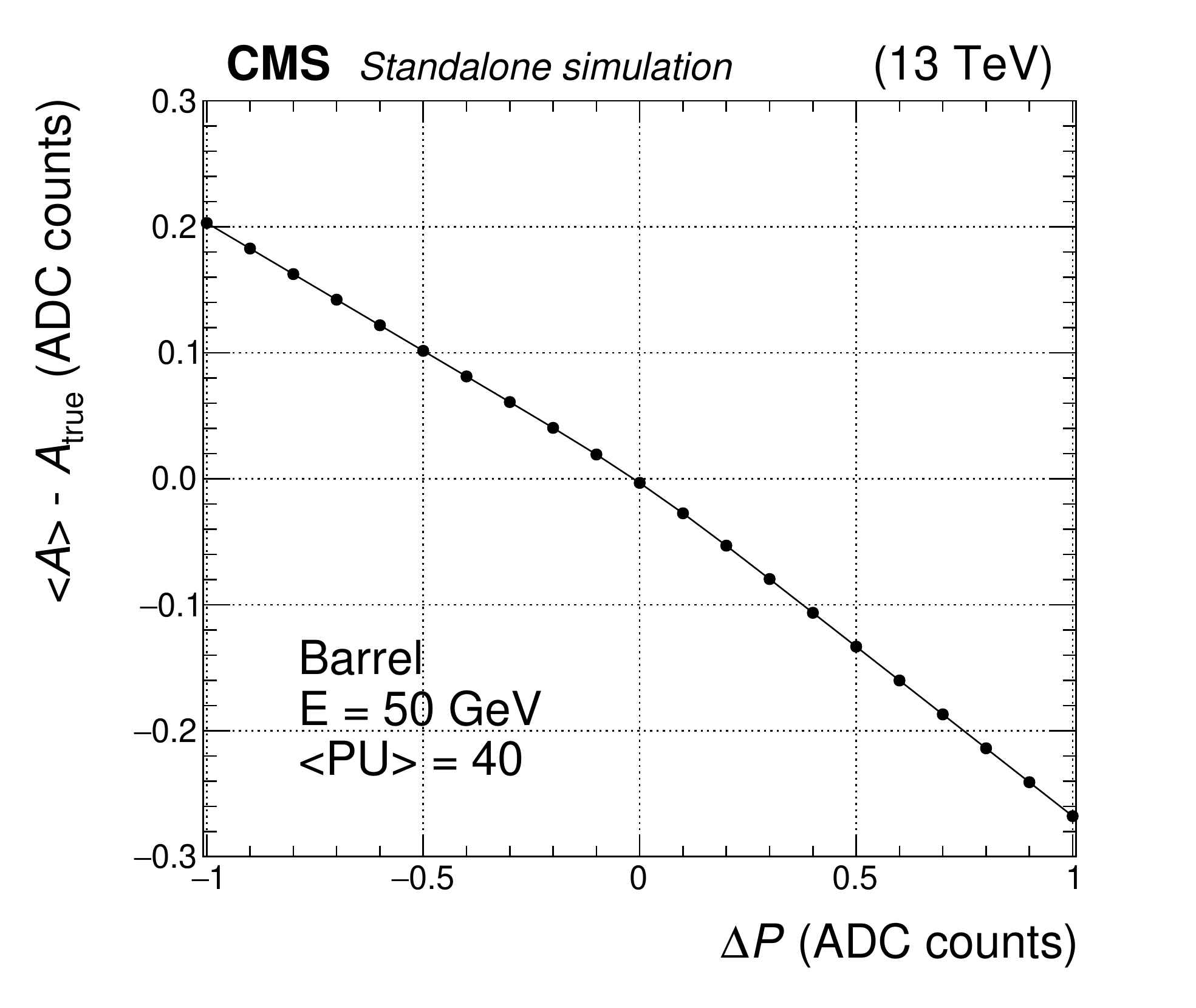}
  \caption{Reconstructed amplitude bias for the IT amplitude,
  $\langle A \rangle-A_\text{true}$, as a function of pedestal shifts
  $\Delta P$, for a single-crystal pulse of $E = 50\GeV$ in the
  EB.\label{fig:pedbias}}

\end{figure}

Figure~\ref{fig:pedbias} shows the absolute amplitude bias for pulses
corresponding to a 50\GeV energy deposit ($E$) in one crystal in the
barrel, as a function of the pedestal baseline shift. The dependence
for deposits in the endcaps is the same. A shift of $\pm$1\unit{ADC}
count produces an amplitude bias up to 0.3\unit{ADC} counts in a
single crystal, corresponding, in the barrel, to an energy-equivalent
shift of about 300\MeV in a $5{\times}5$ crystal matrix. Since the
drift of the pedestal baseline with time can be as much as 2\unit{ADC}
counts in one year of data taking, as shown in Fig.~\ref{fig:pedhisto}
(left), and is coherent in all crystals, the induced bias is
significant, in the range $\approx$(0.5--1)\%, even in the typical
energy range of decay products of the $\PW$, $\PZ$, and Higgs bosons.
Therefore, it is important to monitor and periodically correct the
pedestals in the reconstruction inputs.

The IT amplitude resulting from the $\chi^2$ minimization of
Eq.~(\ref{eqn:chi2}) is also more sensitive to a shift in the position
of the maximum, $T_\text{max}$ of the signal pulse, compared to that
obtained from the weights method~\cite{Bruneliere:2006ra}.  This
timing shift can be caused by variations of the pulse shapes over
time, both independently from crystal to crystal and coherently, as
discussed in Section~\ref{sec:templates}.  A difference in the pulse
maximum position between the measured signal pulse and the binned
template will be absorbed into the $\chi^2$ as nonzero OOT amplitudes,
$A_j$, with $j\neq5$.

\begin{figure}[!htbp]
\centering
\includegraphics[width=0.49\textwidth]{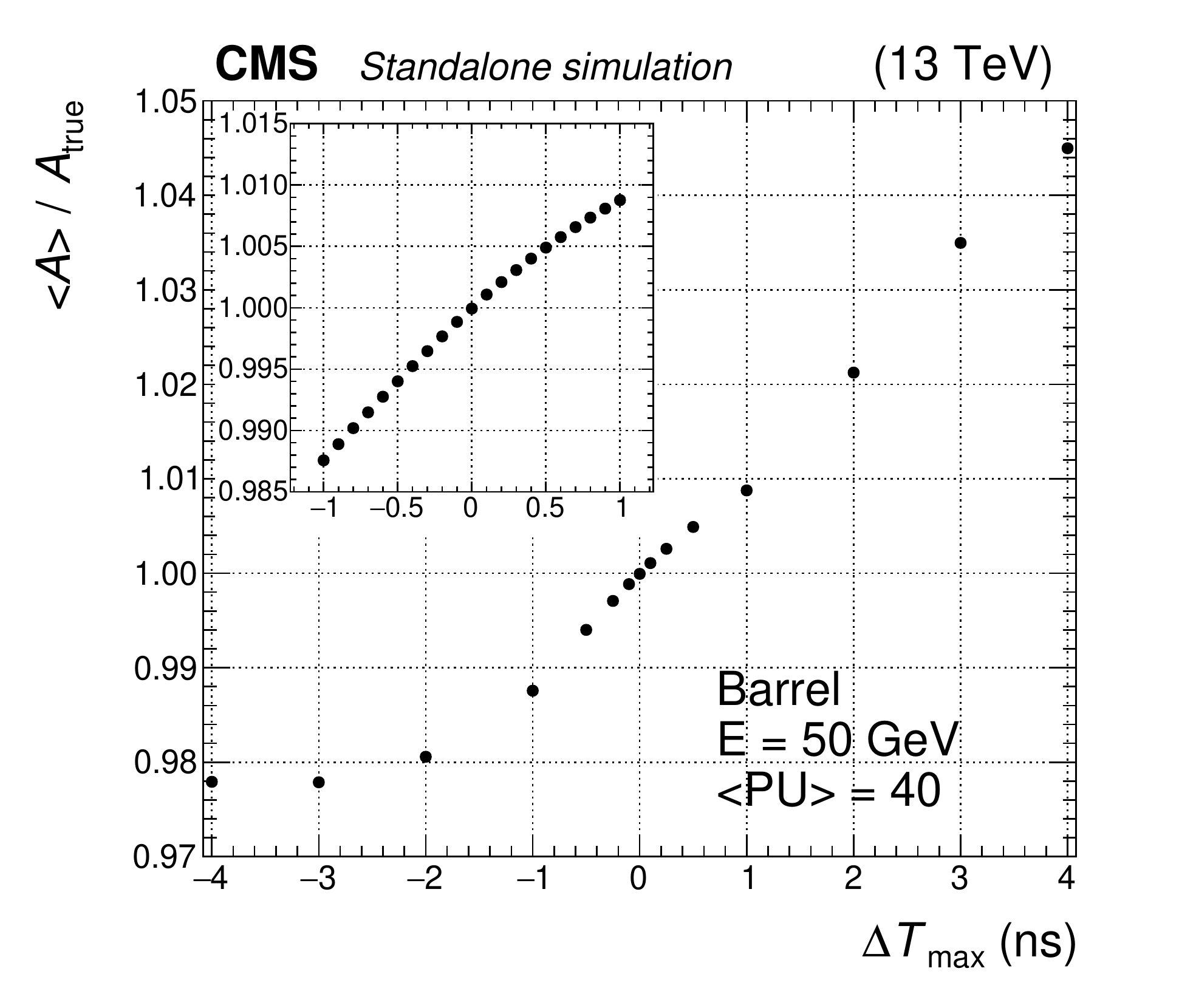}
\includegraphics[width=0.49\textwidth]{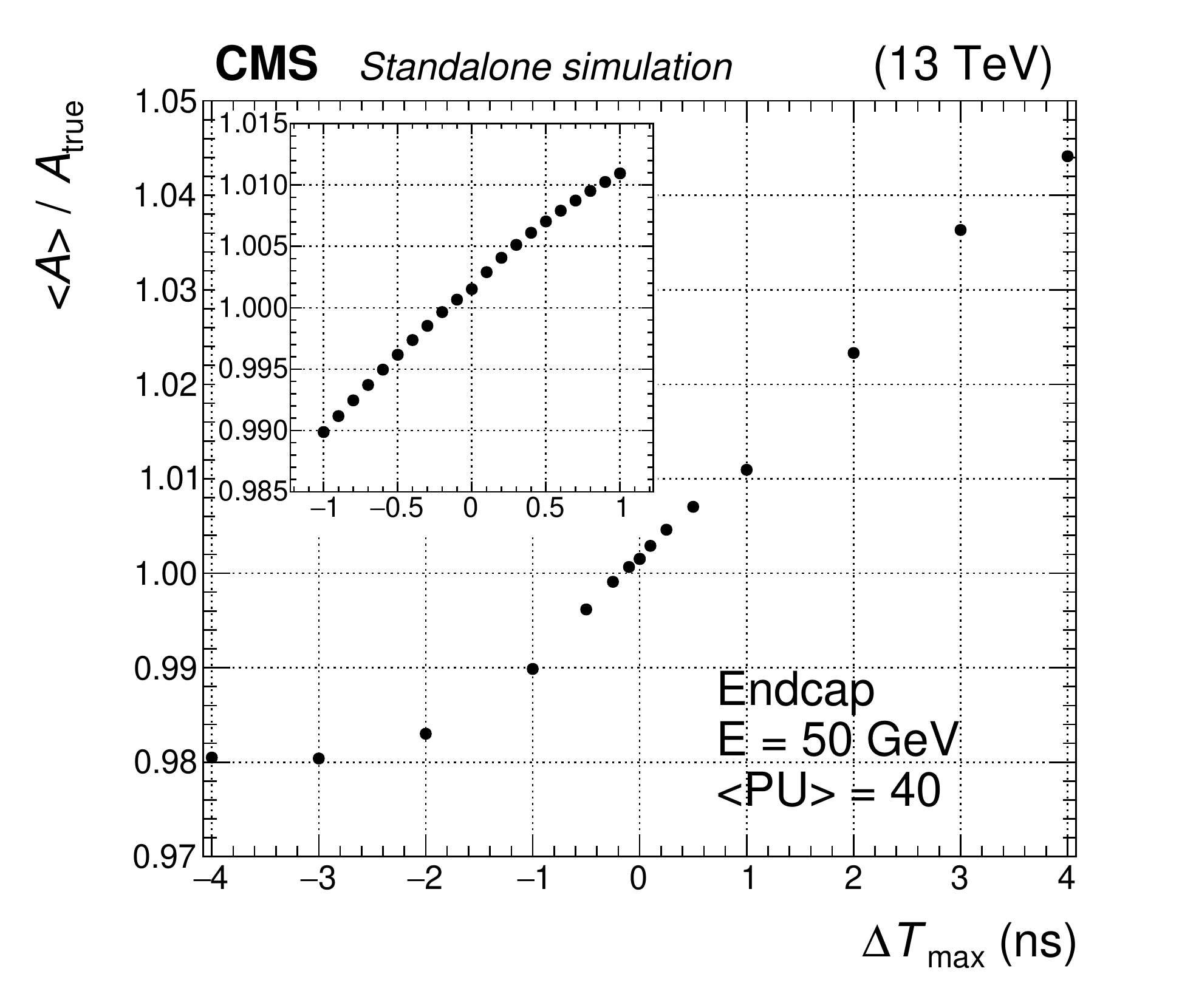}
\caption{Reconstructed amplitude over true amplitude, $\langle A \rangle/A_\text{true}$, as
a function of the timing shift of the pulse template, $\Delta
T_\text{max}$, for a single-crystal pulse of $E=50\GeV$ in the EB
(left) and EE (right).  The insets show an enlargement in the $\pm
1$\unit{ns} range with a finer $\Delta T_\text{max}$
granularity. \label{fig:timebias}}

\end{figure}

To estimate the sensitivity of the reconstructed amplitude to changes
in the template timing $\Delta T_\text{max}$, the amplitude of a given
pulse is reconstructed several times, with increasing values of
$\Delta T_\text{max}$. The observed changes in the ratio of the
reconstructed amplitude to the true amplitude, $\langle
A\rangle/A_\text{true}$, as a function of $\Delta T_\text{max}$, for
single-crystal pulses of 50\GeV in the EB and EE, are shown in
Fig.~\ref{fig:timebias} (left) and (right), respectively. The
difference in shape for positive and negative time shifts is related
to the asymmetry of the pulse shape with respect to the maximum:
spurious OOT amplitudes can be fitted more accurately using the time
samples preceding the rising edge, where pedestal-only samples are
expected, compared to using those on the falling tail.  For positive
$\Delta T_\text{max}$, the net change is positive because the effect
of an increase in the IT contribution is larger than the decrease in
the signal amplitude caused by the misalignment of the template. The
change in reconstructed amplitude at a given $\Delta T_\text{max}$ is
similar for the barrel and the endcaps. Small differences arise mostly
from the slightly different rise time of the barrel and endcap pulses
and the difference in energy distributions from PU interactions in a
single crystal in the two regions. For the endcaps, the residual
offset of $\approx$0.2\% for $\Delta T_\text{max}=0$ has two sources.
First, the larger occupancy of OOT pileup amplitudes per channel
contributes energy coherently to all of the samples within the readout
window. Second, the higher electronics noise leads to a looser
amplitude constraint in the $\chi^2$ minimization of
Eq.~\ref{eqn:chi2}, allowing a larger amplitude to be fitted.  This
offset is reabsorbed in the subsequent absolute energy calibration and
it does not affect the energy resolution.

\begin{figure}[!t]
\centering
\includegraphics[width=0.9\textwidth]{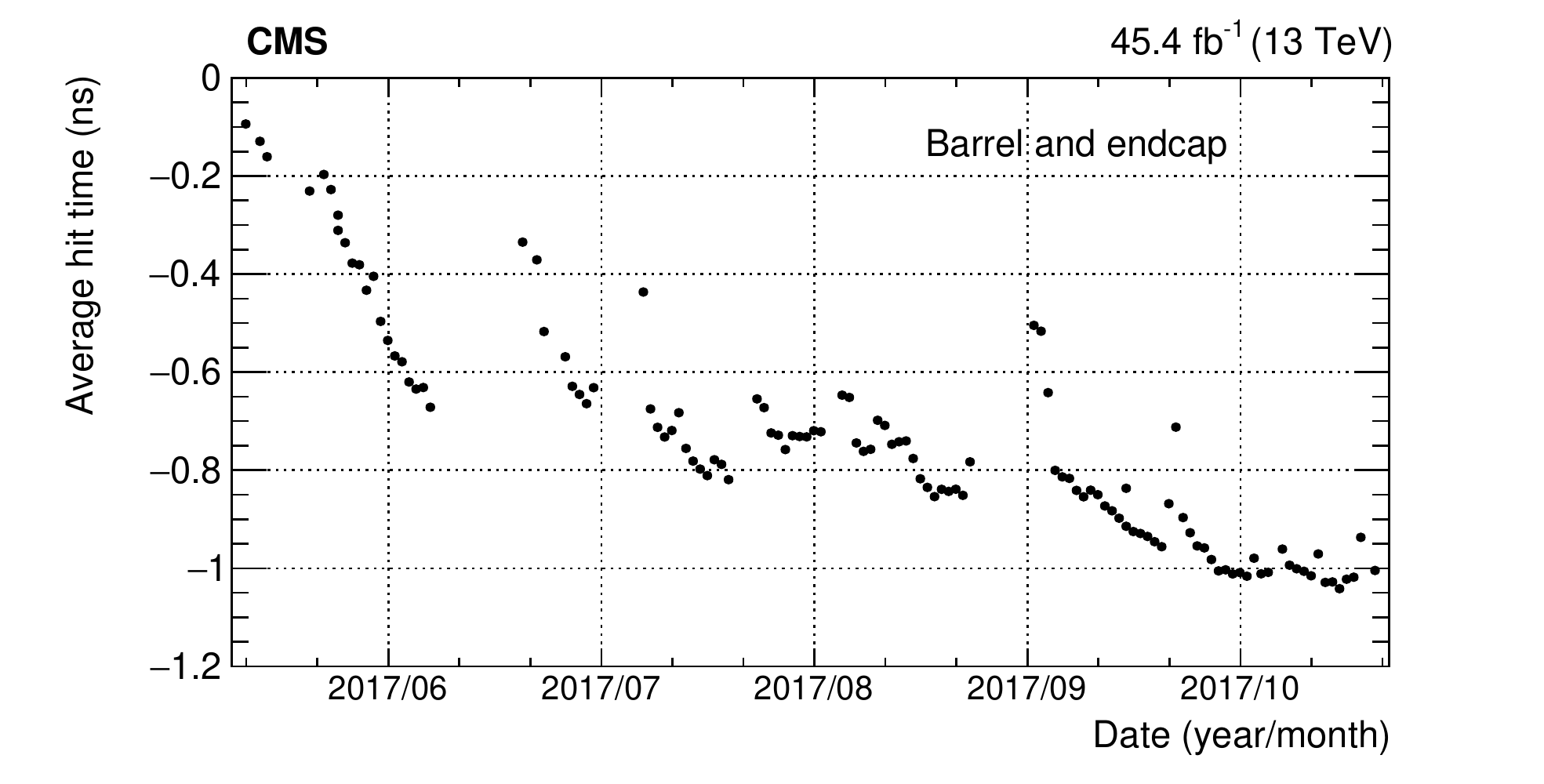}
\caption{Average timing of ECAL pulses in proton-proton collisions
collected in 2017, as measured in Ref.~\cite{Chatrchyan:2009aj}. For
each point, the average of the hits reconstructed in all barrel and
endcaps channels is used. The sharp changes in $T_\text{max}$
correspond to restarts of data taking following LHC technical stops,
as discussed in the text. At the beginning of the yearly data taking,
the timing is calibrated so that the average
$T_\text{max}=0$. \label{fig:avgtiming} }

\end{figure}

The effects of small channel-dependent differences between actual
pulse shapes and the assumed templates are absorbed by the
crystal-to-crystal energy intercalibrations. However, any changes with
time in the relative position of the template will affect the
reconstructed amplitudes, worsening the energy resolution.  This
implies the need to monitor $T_\text{max}$ and periodically correct
the templates for any observed drifts.  The average correlated drift
of $T_\text{max}$ was constantly monitored throughout Run~2, measured
with the algorithm of Ref.~\cite{Chatrchyan:2009aj}.  Its evolution
during 2017 is shown in Fig.~\ref{fig:avgtiming}.  The coherent
variation can be up to 1\unit{ns}. The repeated sharp changes in
$T_\text{max}$ occur when data taking is resumed after a technical
stop of the LHC. They are caused by a partial recovery in crystal
transparency while the beam is off, followed by a rapid return to the
previous value when irradiation resumes. A similar trend was measured
in the other years of data-taking during Run~2.

The measured time variation is crystal dependent, since the integrated
radiation dose depends on the crystal position, and since there are
small differences in the effect between crystals at the same
$\eta$. For this reason the pulse templates are measured in situ
multiple times during periods with collision data, and a specific
pulse template is used for each channel.  The measurement described in
Section~\ref{sec:templates} is repeated after every LHC technical
stop, when a change of the templates is expected because of partial
recovery of the crystal transparency, or when the $\abs{\Delta
T_\text{max}}$ was larger than 250\unit{ps}.

\section{Performance with simulations and collision data}
\label{sec:results}

In this section, the performance of the ECAL local reconstruction with
the multifit algorithm is compared with the weights
method~\cite{Bruneliere:2006ra}.  Simulated events with a PU
typical of Run~2 (a Poisson distribution with a mean of 40) and
collision data collected in 2016--2018 are used. The data comparisons
are performed for low-energy photons from $\PGpz\to\PGg\PGg$
decays, and for high-energy electrons from $\PZ\to\Pep\Pem$ decays.

\subsection{Suppression of out-of-time pileup signals}
\label{sec:singlecrystal}

The motivation for implementing the multifit reconstruction is to
suppress the OOT pileup energy contribution, while reconstructing IT
amplitudes as accurately as possible. To show how well the multifit
reconstruction performs, the resolution of the estimated IT energy is
compared for single crystals, as a function of the average number of
PU interactions. This study was performed using simple
pseudo-experiments, where the pulse shape is generated according to
the measured template for a barrel crystal at $\abs{\eta}\approx0$.
The appropriate electronics noise, equal to the average value measured
in Run~2, together with its covariance matrix, is included. The effect
of the PU is simulated assuming that the number of additional
interactions has a Poisson distribution about the mean expected value
and that these interactions have an energy distribution corresponding
to that expected for minimum bias events at the particular value of
$\eta$ of the crystal. The pseudo-experiments are performed for two
fixed single-crystal energies: 2 and 50\GeV. For a single crystal, the
amplitude is related directly to the energy only through a constant
calibration factor, thus the resolution of the uncalibrated amplitude
equals the energy resolution.  The resolution of a cluster receives
other contributions that may degrade the intrinsic single-crystal
energy measurement precision, such as a nonuniform response across
several crystals, within the calibration uncertainties. These
considerations are outside the scope of this paper.

\begin{figure}[!htbp]
\centering
\includegraphics[width=0.49\textwidth]{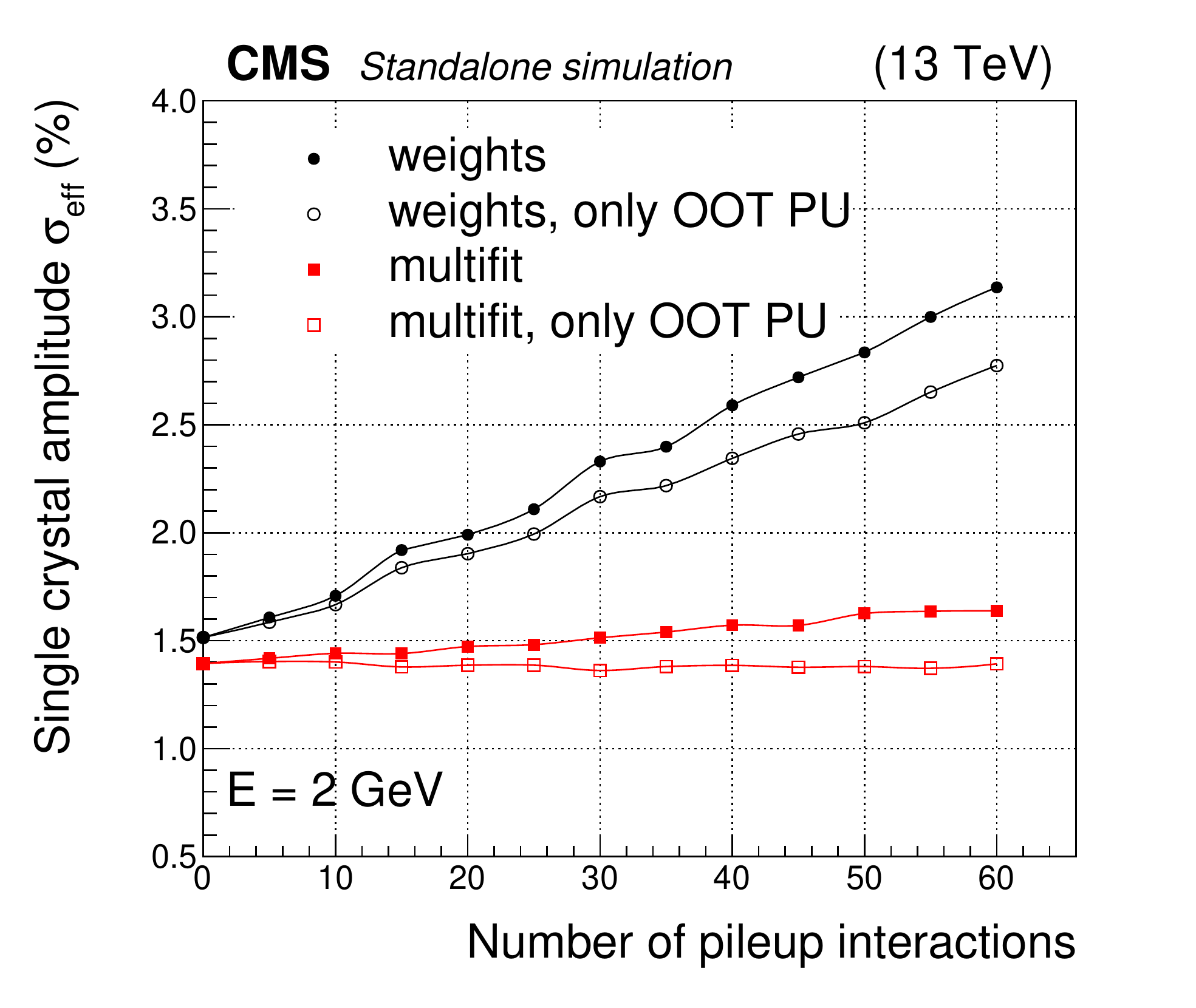}
\includegraphics[width=0.49\textwidth]{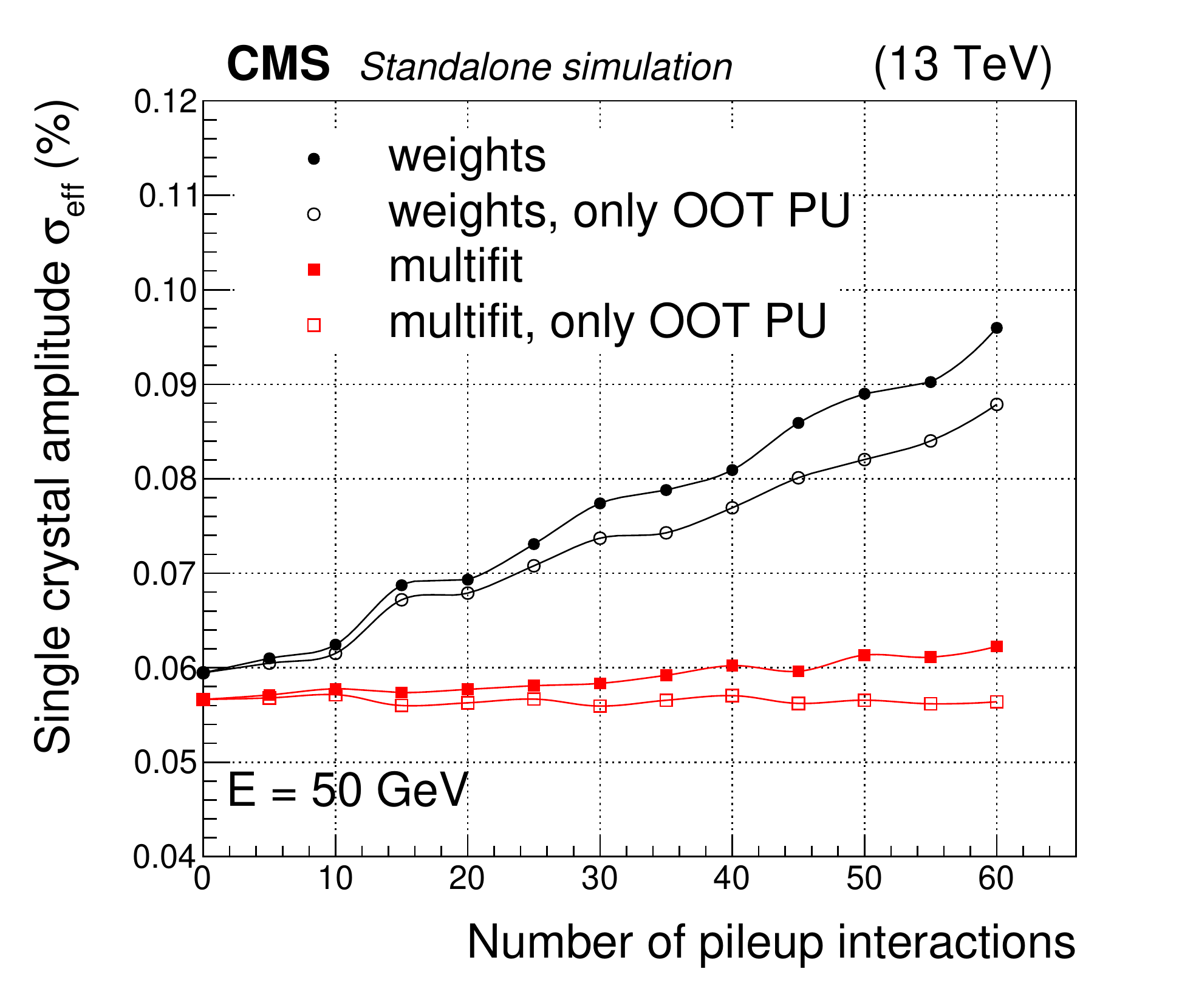}
  \caption{Measured amplitude resolution for two generated energy
  deposits ($E=2\GeV$ or $E=50\GeV$) in a single ECAL barrel crystal,
  at $\eta=0$, reconstructed with either the multifit or the weights
  algorithm. Filled points show the effective resolution expressed as
  the difference between the reconstructed energy and the true energy,
  divided by the true energy. Open points show the percent
  resolution estimated when the true energy is replaced with the sum
  of the true energy and the in-time pileup
  energy. \label{fig:reso_vs_pu}}

\end{figure}

The amplitude resolution is estimated as the effective standard
deviation $\sigma_\text{eff}$, calculated as half of the smallest
symmetrical interval around the peak position containing 68.3\% of the
events.  The PU energy from IT interactions constitutes an
irreducible background for both energy reconstruction methods. It is
expected that event-by-event fluctuations of this component degrade
the energy resolution in both cases as the PU increases.  On the
other hand, the fluctuations in the energy from all the OOT
interactions are suppressed significantly by the multifit algorithm,
in contrast to the situation for the weights reconstruction, where
they contribute further to the energy resolution deterioration at
large average PU.  This is shown in Fig.~\ref{fig:reso_vs_pu}, for
the two energies considered in this study. The reconstructed energy is
compared with either the true generated energy (corrected for both IT
and OOT PU) or the sum of the energy from the IT pileup and the true
energy (corrected only for the effect of OOT PU). In the latter case,
the amplitude resolution for the multifit reconstruction does not
depend on the number of interactions, showing that this algorithm
effectively suppresses the contributions of the OOT PU. The offset in
resolution in the case of no PU between the two methods, in this ideal
case, is due to the improved suppression of the electronic noise
resulting from the use of a fixed pedestal rather than the
event-by-event estimate used in the weights method. In the data,
additional sources of miscalibration may further worsen the energy
resolution. Such effects are considered in the full detector
simulation used for physics analyses, described below, but are not
included in this stand-alone simulation.

Simulations performed for an upgraded EB, planned for the
high-luminosity phase of the LHC~\cite{ApollinariG.:2017ojx}, have
shown that the multifit algorithm can subtract OOT PU for energies
down to the level of the electronic noise, for
$\sigma_\text{noise}>10\MeV$, for PU values up to 200 with 25\unit{ns}
bunch spacing. This future reconstruction method will benefit from a
more frequent sampling of the pulse shape, at 160 MHz, and from a
narrower signal pulse to be achieved with the upgraded front-end
electronics~\cite{phase2ecalTDR}.

\subsection{Energy reconstruction with simulated data}
\label{sec:resultsfullsim}

\begin{figure}[!htbp]
\centering
\includegraphics[width=0.49\textwidth]{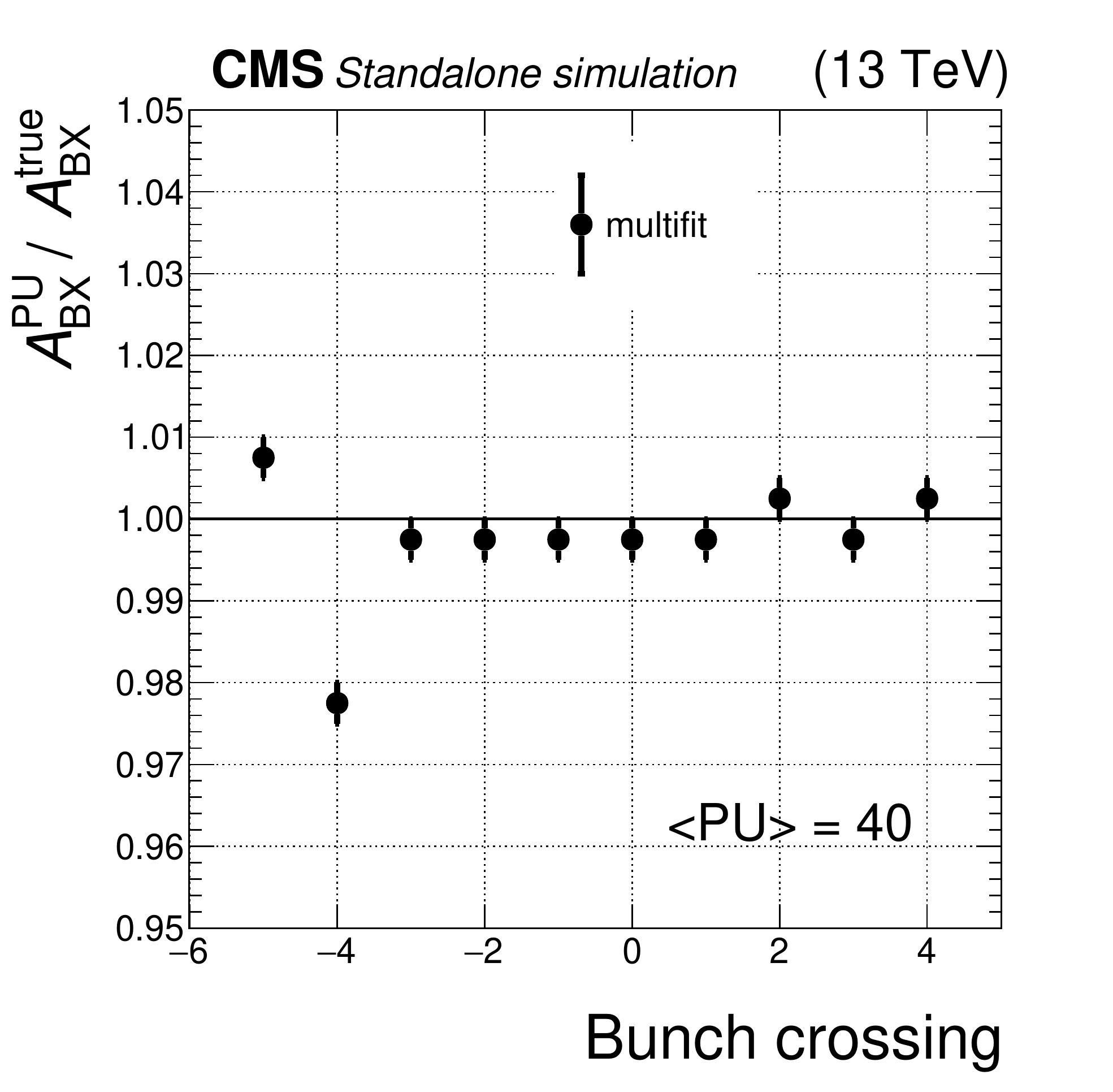}
\includegraphics[width=0.49\textwidth]{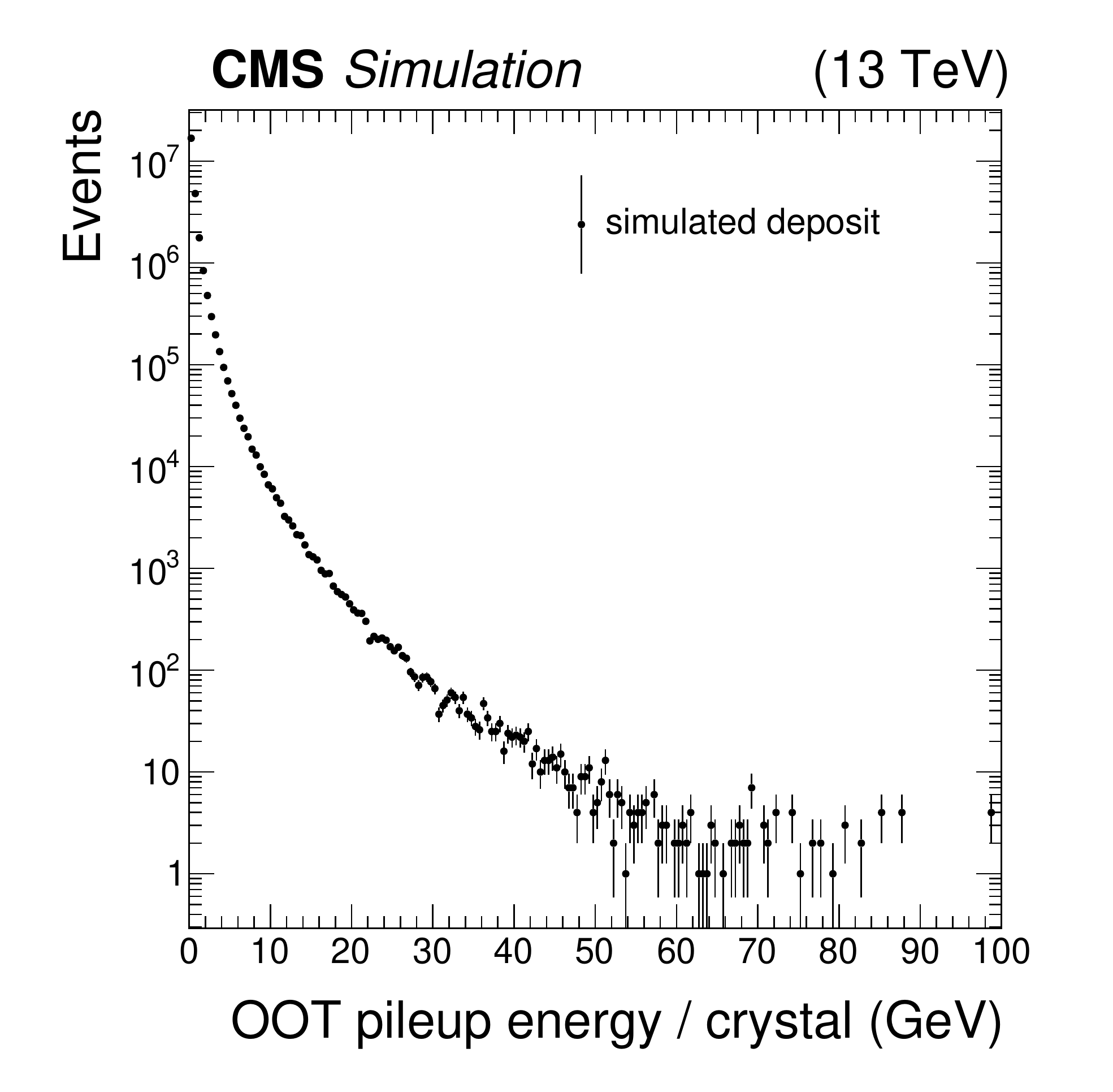}
  \caption{Left: bias in the out-of-time amplitude estimated by the
  multifit algorithm as a function of BX, for the bunch crossings
  $-5\le\mathrm{BX}\le+4$. The in-time interaction corresponds to BX
  $=$ 0 in the figure. The bias is estimated as the mode of the
  distribution of the ratio between the measured and the true
  energy. Only statistical uncertainties are shown.  Right: energy
  spectrum in an ECAL barrel crystal, at
  $\eta\approx0$. \label{fig:ootbias}}

\end{figure}

The ability of the multifit algorithm to estimate the OOT amplitudes
and, consequently, to estimate the IT amplitude is demonstrated in
Fig.~\ref{fig:ootbias} (left). Simulated events are generated with an
average of 40 PU interactions, with an energy spectrum per EB crystal
as shown in Fig.~\ref{fig:ootbias} (right).  The reconstructed energy
assigned by the multifit algorithm to each BX from $-$5 to $+$4 is
compared with the generated value. The IT contribution corresponds to
BX $=$ 0. Amplitudes are included with energy larger than 50\MeV, a
value corresponding approximatively to one standard deviation of the
electronic noise~\cite{Khachatryan:2015iwa}. The mode of the
distribution of the ratio between the reconstructed and true energies
of OOT PU pulses and true energies,
$A^\mathrm{PU}_\mathrm{BX}/A^\text{true}_\mathrm{BX}$, with BX in the
range $[-5,$\ldots$,+4]$, is equal to unity within $\pm 2.5\%$ for all
the BXs.  The OOT interactions simulated in these events cover a range
from 12 BXs before to 3 BXs after the IT interaction, as is done in
the full simulation used in CMS. The distribution of the measured to
true energy becomes asymmetric at the boundaries of the pulse readout
window ($\mathrm{BX}=-5$, $-$4, and $-$3), because the contributions
of earlier interactions cannot be resolved with the information
provided by the 10 digitized samples. However, this does not introduce
a bias in the IT amplitude since the energy contribution from very
early BXs below the maximum of the IT pulse is negligible. The
remaining offset of $\approx$0.2\% in the median of
$A^\mathrm{PU}_\mathrm{BX}/A^\text{true}_\mathrm{BX}$ for BXs close to
zero is due to the requirement that all the $A_j$ values are
nonnegative, \ie, any spuriously fitted OOT pulse can only subtract
part of the in-time amplitude. This offset is absorbed in the absolute
energy scale calibration and does not affect the energy resolution.

The energy from an electromagnetic shower for a high-momentum electron
or photon is deposited in several adjacent ECAL crystals. A clustering
algorithm is required to sum together the deposits of adjacent
channels that are associated with a single electromagnetic shower.
Corrections are applied to rectify the cluster partial containment
effects.  In the present work, we use a simple clustering algorithm
that sums the energy in a $5{\times}5$ crystal matrix centered on the
crystal with the maximum energy deposit. This approach is adequate for
comparing the performance of the two reconstruction algorithms,
especially in regions with low tracker material (\eg,
$\abs{\eta}<0.8$), where the fraction of energy lost by electrons by
bremsstrahlung (and subsequent photon conversions) is small. Here,
more than 95\% of the energy is contained in a $5{\times}5$ matrix.
To reduce the fraction of events with partial cluster containment
caused by early bremsstrahlung and photon conversion, a selection is
applied to the electrons and photons.  In the simulation, events with
photon conversions are rejected using Monte Carlo information, whereas
in data a variable that uses only information from the tracker is
adopted, as described later.

The relative performance of the two reconstruction algorithms is
evaluated on a simulated sample of single-photon events generated
by \GEANTfour with a uniform distribution in $\eta$ and a flat
transverse momentum $\pt$ spectrum extending from 1 to 100\GeV.  The
photons not undergoing a conversion before the ECAL surface are
selected by excluding those that match geometrically electron-positron
pair tracks from conversions in the simulation. For the retained
photons, the energy is mostly contained in a $5{\times}5$ matrix of
crystals, and no additional corrections are applied.

\begin{figure}[!htbp]
\centering
\includegraphics[width=0.49\textwidth]{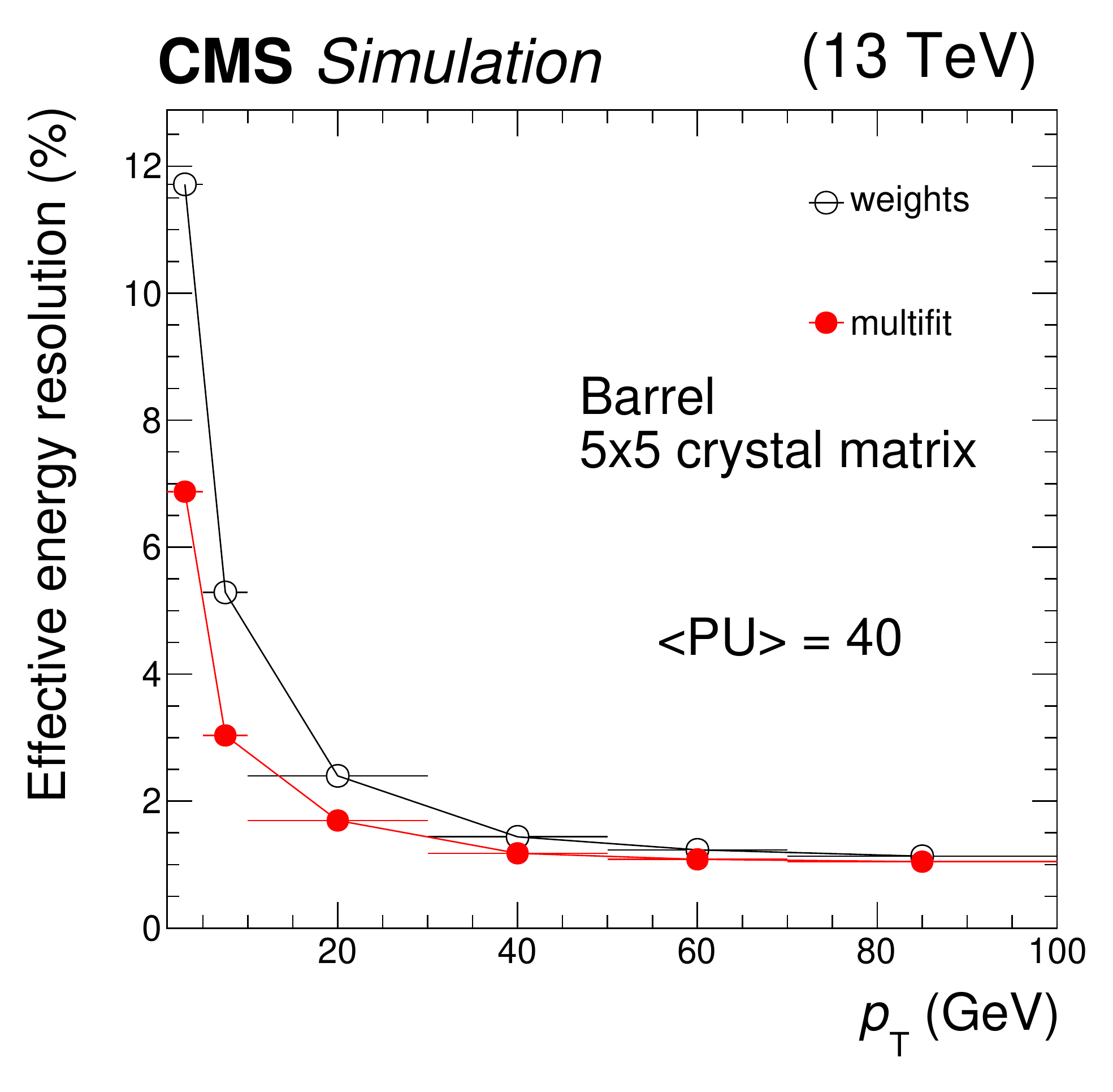}
\includegraphics[width=0.49\textwidth]{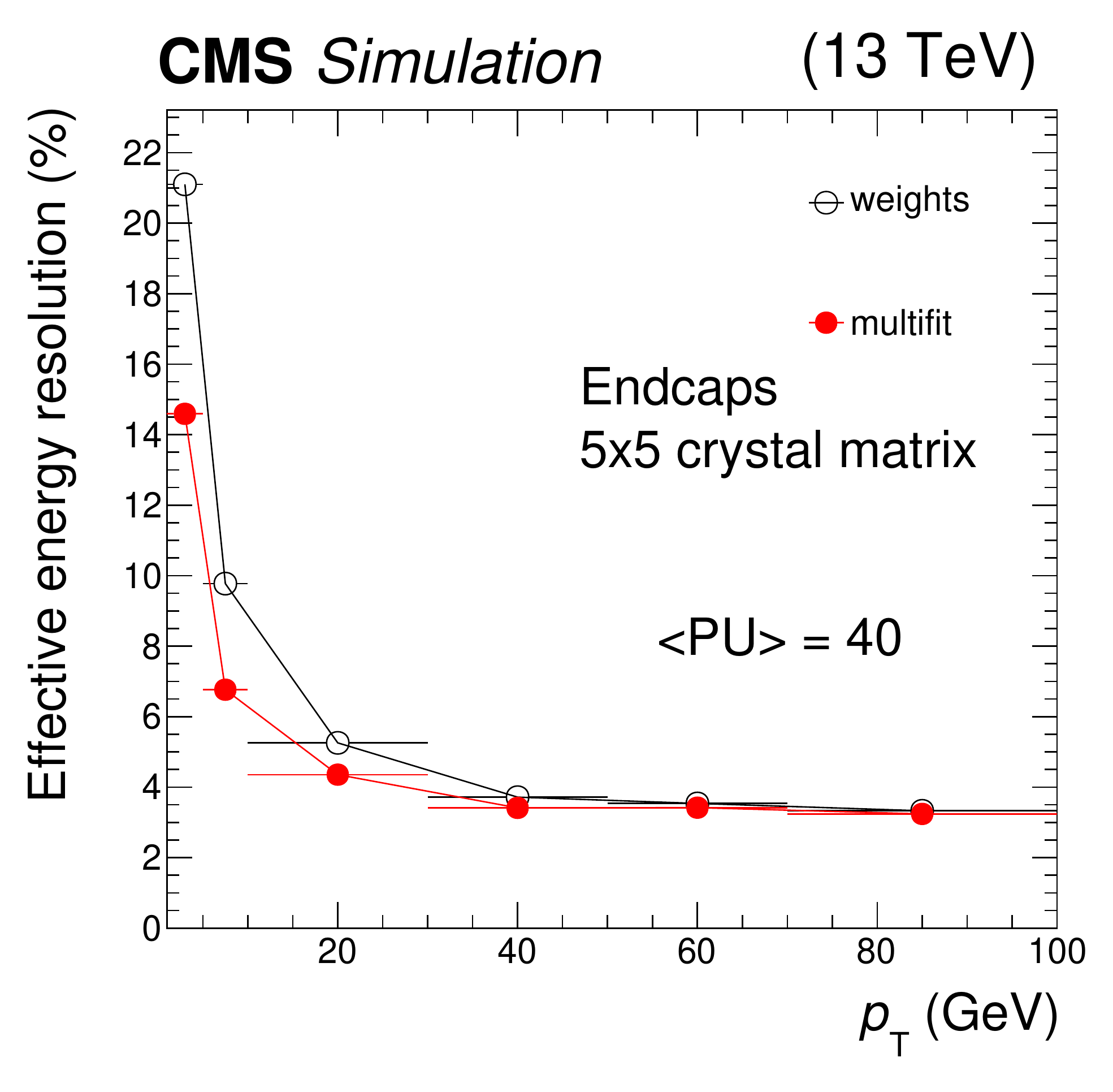}
  \caption{Effective energy resolutions for nonconverted photons in
barrel (left) and endcaps (right) as a function of the generated
\pt of the photon. The photons are generated with a
uniform \pt distribution and their interaction is obtained with the
full detector simulation. The average number of PU interactions is
40. The horizontal error bars represent the bin width. The statistical
uncertainties are too small to be displayed. \label{fig:resolution}}

\end{figure}

The ratio between the reconstructed energy in the $5{\times}5$ crystal
matrix and the generated photon energy,
$E_{5{\times}5}/E_\text{true}$, for nonconverted photons with a
uniform distribution in the range $1<\pt^\text{true}<100\GeV$ is
histogramed.  For both reconstruction algorithms, the distributions
show a non-Gaussian tail towards lower values, caused by the energy
leakage out of the $5{\times}5$ crystal matrix, which is not corrected
for. To account for this, $\sigma_\text{eff}$, as defined in
Section~\ref{sec:singlecrystal}, is used to quantify the energy
resolution.  The average energy scale of the reconstructed clusters is
shifted downwards for the multifit method, whereas it is approximately
unity for the weights reconstruction. As stated earlier, this is
because the amplitudes for the OOT pulses ($A_j$ with $j \neq 5$) are
constrained to be positive. In the reconstruction of photons used by
CMS such a shift is corrected for, a posteriori, by a dedicated
multivariate regression, which simultaneously corrects the residual
dependence of the energy scale on the cluster containment and IT
pileup.  This correction is applied in the HLT and, with a more
refined algorithm, in the offline event reconstruction.  This type of
cluster containment correction was developed in
Run~1~\cite{Khachatryan:2015hwa,Khachatryan:2015iwa} and has been used
subsequently. In this approach, the shift of the
$E_{5{\times}5}/E_\text{true}$ distribution is corrected by rescaling
the resolution estimator, $\sigma_\text{eff}$, by $m$, estimated as
the mean of a Gaussian function fitting the bulk of the distribution,
and expressed in percent.  The variation of $\sigma_\text{eff}$ as a
function of the true \pt of the photon, is shown in
Fig.~\ref{fig:resolution}.

The improvement in the precision of the energy measurement is
significant for the full range of $\pt$ considered. Expressed as a
quadratic contribution to the total, it varies from 10 (15)\% in the
barrel (endcaps) for photons with $\pt<5\GeV$, to 0.5 (1.0)\% at
$\pt=100\GeV$. The improvement is larger at low $\pt$, since the
relative contribution of the energy deposits from PU interactions,
which have the characteristic momentum spectrum shown in
Fig.~\ref{fig:ootbias} (right), is relatively larger. This is
particularly relevant for suppressing the PU contribution to low-$\pt$
particles that enter the reconstruction of jets and missing transverse
momentum with the particle-flow algorithm used in
CMS~\cite{Sirunyan:2017ulk}, thus preserving the resolution achieved
during
Run~1~\cite{Khachatryan:2016kdb,CMS-DP-2018-028,Sirunyan:2019kia}. The
improvement grows with $\abs{\eta}$ both within the EB and within the
EE, because of the increasing probability of overlapping pulses from
PU.  The improvement is larger in the barrel, even though the PU
contribution is smaller than in the endcaps, because the lower
electronic noise allows a more stringent constraint of the amplitudes
in the multifit.  For photons, the improvement extends above
$\pt\approx50\GeV$, because of the higher number of digitized samples
of the pulse shape used, and the suppression of the residual OOT PU
contribution. The energy resolution becomes constant at very high
energies, above a few hundred \GeV, where it is dominated by sources
other than the relatively tiny contribution of OOT pileup energy, such
as nonuniformities in the energy response of different crystals
belonging to the same cluster. The improvement in energy resolution is
also expected to be valid for electrons with $\pt>20 (10)$\GeV in the
barrel (endcaps), since the electron momentum resolution is dominated
by the ECAL cluster measurement above these \pt
values~\cite{Khachatryan:2015hwa}.

\subsection{Energy reconstruction with Run~2 data}
\label{sec:resultsdata}

\subsubsection{Effect on low energy deposits using $\Pgpz\to\gamma\gamma$}
\label{sec:pizero}

\begin{figure}[!htbp]
\centering
\includegraphics[width=0.49\textwidth]{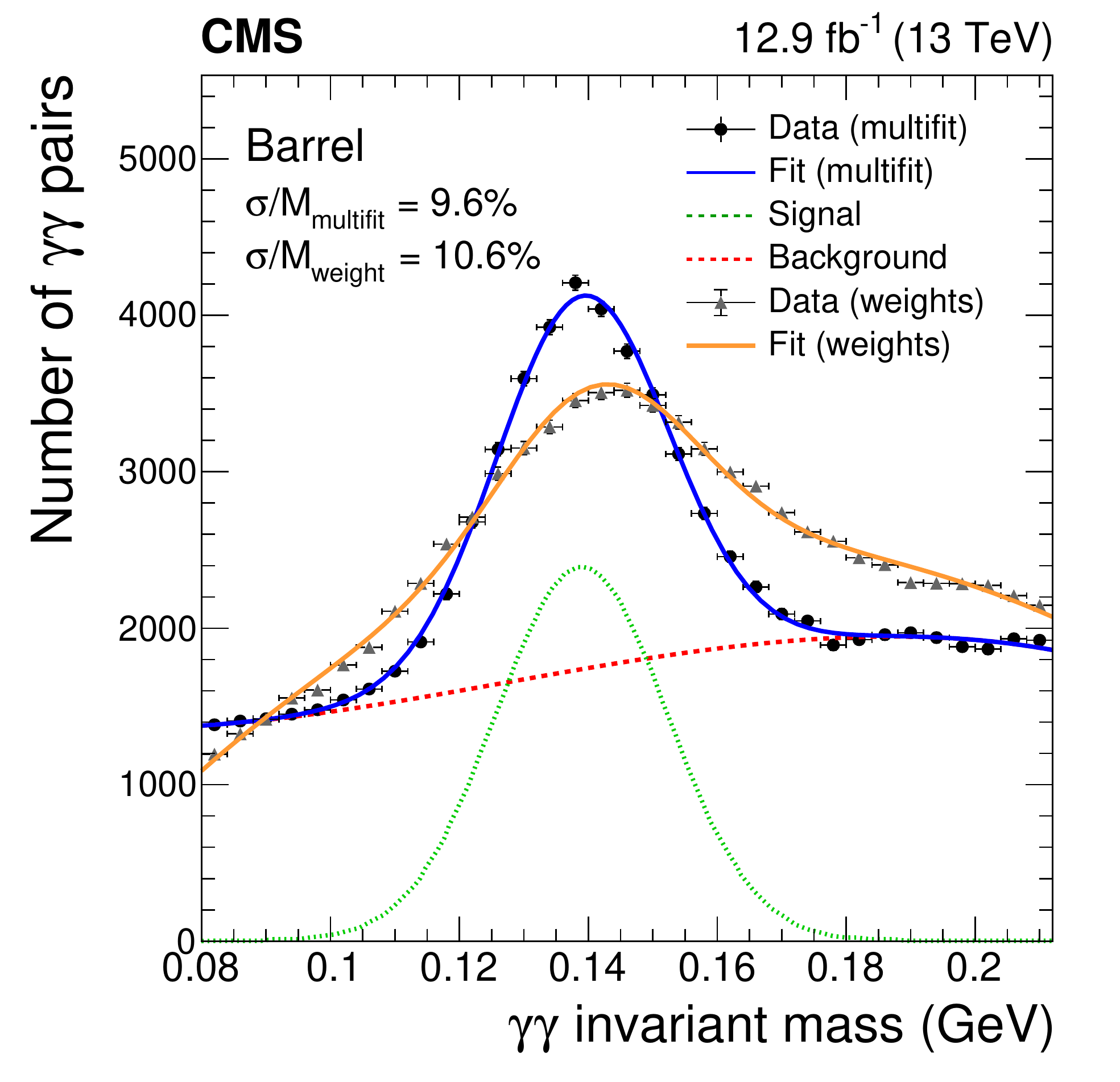}
\includegraphics[width=0.49\textwidth]{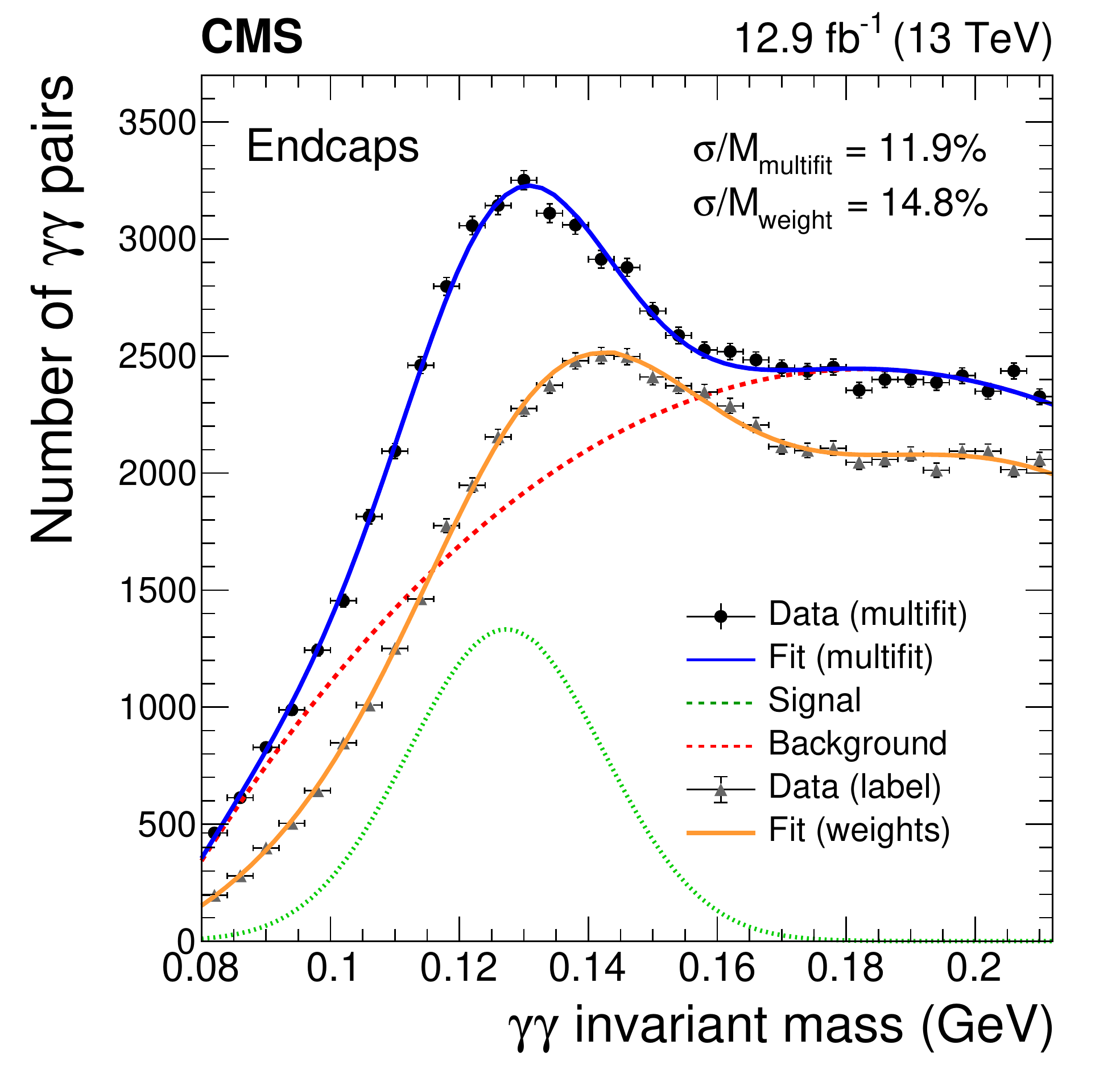}
  \caption{The invariant mass distribution of the two photons for the
 selected $\PGpz\to\gamma\gamma$ candidates in the barrel (left) and
 endcaps (right), for the single-crystal amplitudes measured with
 either the weights or the multifit reconstruction. A portion of
 collision data with typical Run~2 conditions, recorded during July
 2018, is used. Vertical error bars represent the statistical
 uncertainty.  The result of the fit with a Gaussian distribution
 (green dotted line) plus a polynomial function (red dashed line) is
 superimposed on the measured distributions for the multifit case
 (dark blue solid line).  For the weights case the same model is used,
 but only the total likelihood is shown superimposed (light orange
 solid line).\label{fig:pi0}}
\end{figure}

The improvement in the energy resolution for low-energy clusters is
quantified in data using $\PGpz$ mesons decaying into two photons. The
$\pt$ spectrum of the photons, selected by a dedicated calibration
trigger~\cite{Chatrchyan:2013dga}, falls very fast and most of the
photons have a \pt in the range of 1--2\GeV.  The photon energy in
this case is reconstructed summing the energy of the crystals in a
$3{\times}3$ matrix.  Figure~\ref{fig:pi0} shows the diphoton
invariant masses when both clusters are in the EB (left) and when both
are in EE (right). The invariant mass distributions obtained with the
weights and the multifit methods are compared, using a subset of the
$\PGpz$ calibration data collected during 2018. The position of the
peak, $M$, is affected by OOT PU differently in the multifit method
and in the weights algorithm.
Since the $\PGpz\to\PGg\PGg$ process is only used to calibrate the
relative response of a crystal with respect to others, the absolute
energy scale is not important here. The energy scale is determined
separately by comparing the position of the $\PZ\to\Pep\Pem$ mass peak
in data and simulation. On the other hand, the improvement in mass
resolution, $\sigma/M$, is significant, 4.5\% (8.8\%) in quadrature in
the barrel (endcaps).

\begin{figure}[!htbp]
\centering
\includegraphics[width=0.49\textwidth]{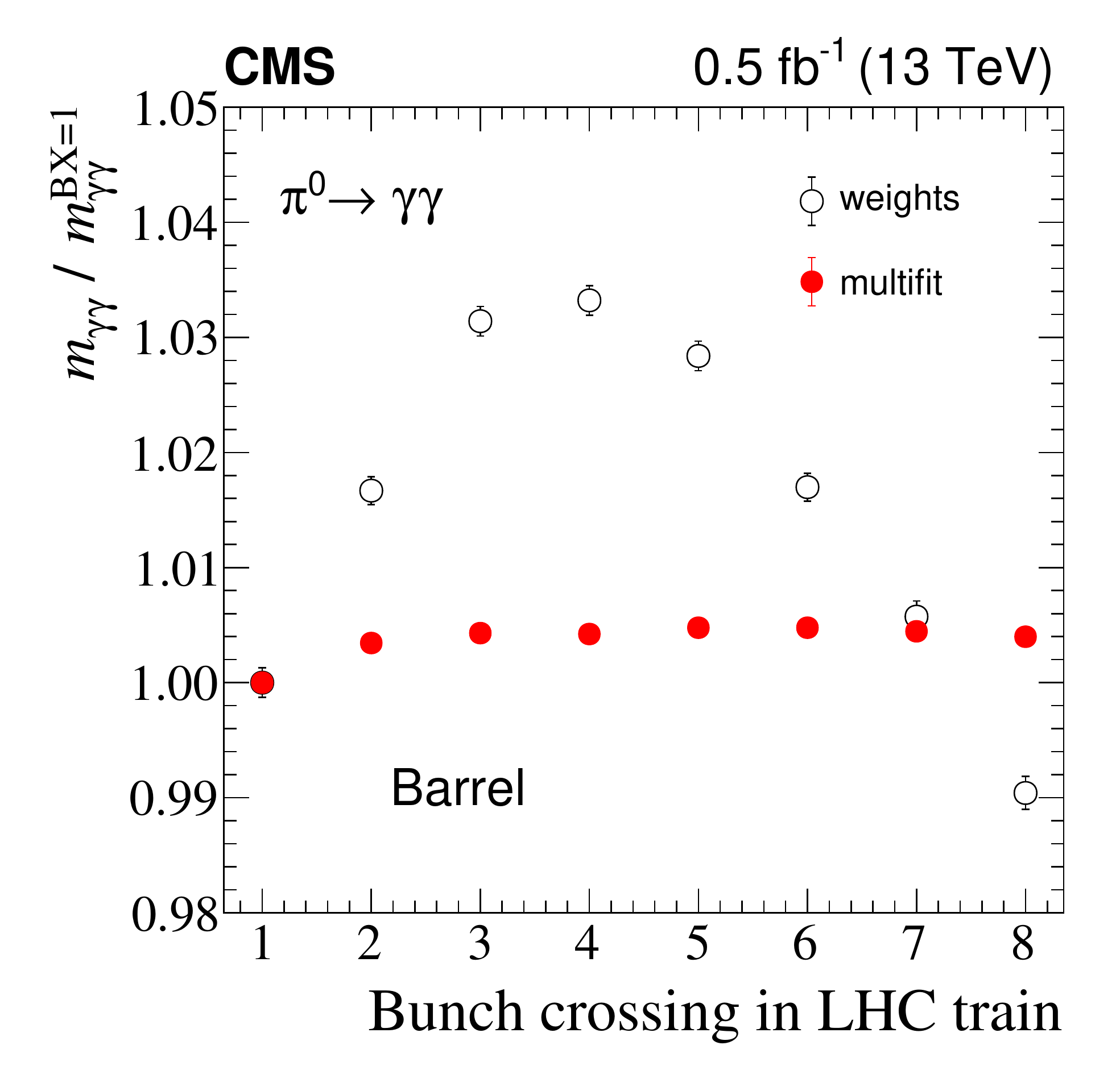}
\includegraphics[width=0.49\textwidth]{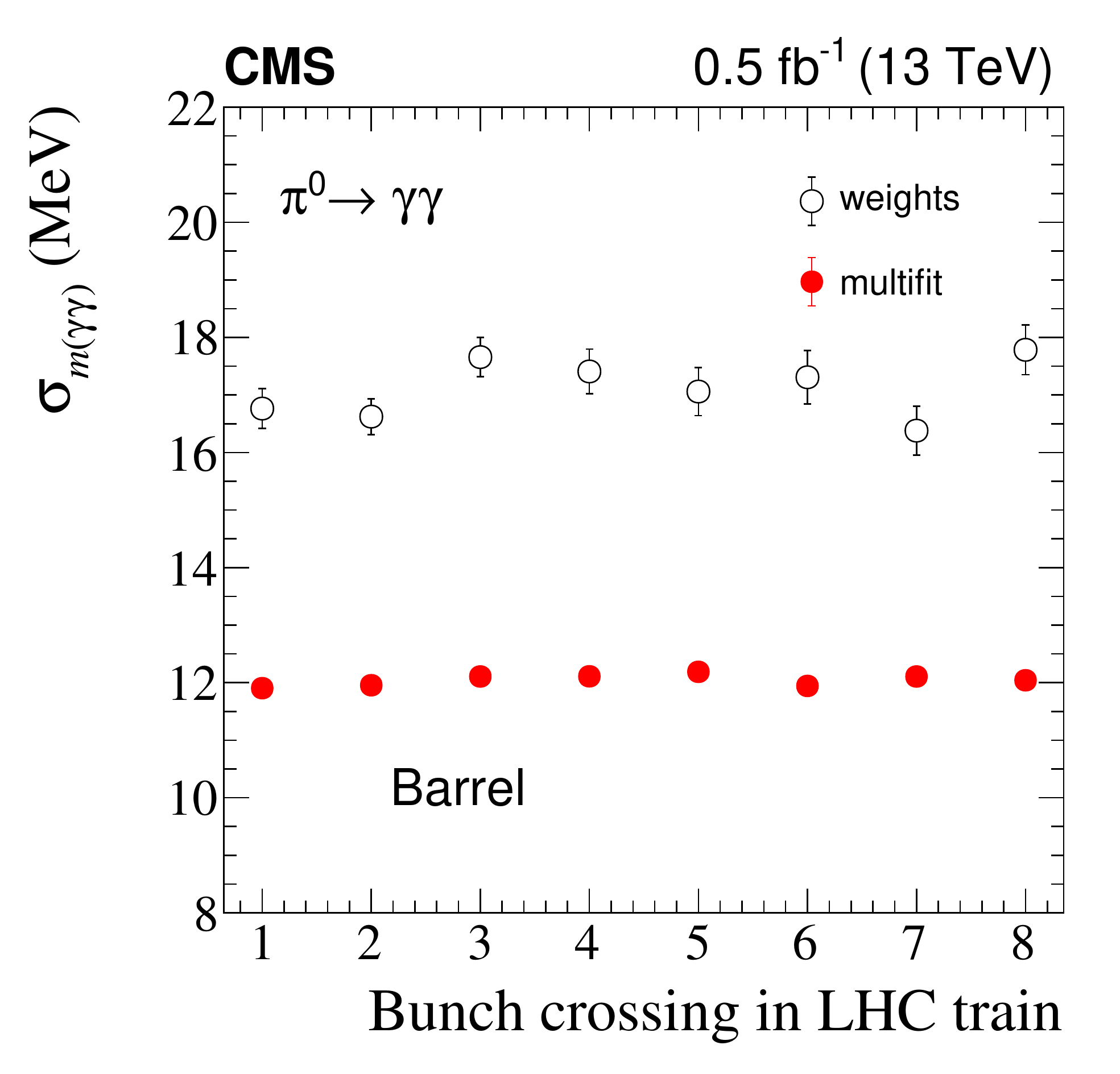}
  \caption{Peak position, normalized to the mass measured in the first
  BX of the train, (left) and Gaussian resolution
  $\sigma_{m({\PGg\PGg})}$ (right) of the invariant mass
  distribution of $\PGpz\to\PGg\PGg$ decays with both photons in
  the EB, within a bunch train of 8 colliding bunches from an LHC fill
  in October 2017. Error bars represent the statistical
  uncertainty. The single-crystal energy is reconstructed either with
  the weights method (open circles) or with the multifit method
  (filled circles). Each point is obtained by fitting the diphoton
  invariant mass distribution in collisions selected from a single BX
  of the train. \label{fig:pi0bx}}
\end{figure}

At the end of 2017, the LHC operated for a period of about 1 month
with a filling scheme with trains of 8 bunches alternated with 4 empty
BXs.  The resilience of the multifit method to OOT pileup had a
particularly positive effect in this period, since the bunch-to-bunch
variations in OOT PU are larger than with the standard LHC filling
schemes used in Run~2. All the bunches of a given train provide
approximately the same luminosity, about $5.5{\times}10^{27}\percms$,
so the average number of PU interactions is the typical one of Run~2
(about 34, with peaks up to 80).  Data from this period is used to
assess the sensitivity of the algorithms to OOT interactions by
estimating the invariant mass peak position of the $\PGpz$ mesons as a
function of BX within each LHC bunch train.  The measured invariant
mass, normalized to that measured in the first BX of the train, is
shown in Fig.~\ref{fig:pi0bx} (left). The peak position, estimated
with the weights algorithm, increases for BXs towards the middle of
the bunch train, where the contribution from OOT collisions is larger,
and then decreases again towards the end of the train. In contrast,
for the multifit reconstruction, the peak position remains stable
within $\pm 0.4\%$ with respect to the value observed in the first BX
of the train. The overall resolution in the diphoton invariant mass
improves significantly using the multifit algorithm, and, within the
precision of the measurement, is insensitive to the variations of OOT
PU for different BX within the train. This is shown in
Fig.~\ref{fig:pi0bx} (right).

\subsubsection{Effect on high energy deposits using $\cPZ\to\Pep\Pem$}
\label{sec:zee}

The performance of the two algorithms for high-energy electromagnetic
deposits is estimated using electrons from $\PZ\to\Pep\Pem$ decays.
Electrons with $\pt>25\GeV$ are identified with tight electron
identification criteria, using a discriminant based on a multivariate
approach~\cite{Khachatryan:2015hwa}. To decouple the effects of
cluster containment corrections from the single-crystal resolution,
$5{\times}5$ crystal matrices are used to form clusters.  The sample
is enriched in low-bremsstrahlung electrons by selecting with an
observable using only tracker information, $f_\text{brem}$, which
represents the fraction of momentum, estimated from the track, lost
before reaching the ECAL. It is defined as
$f_\text{brem}=(p_\text{in}-p_\text{out})/p_\text{in}$, where
$p_\text{in}$ and $p_\text{out}$ are the momenta of the track
extrapolated to the point of closest approach to the beam spot and
estimated from the track at the last sensitive layer of the tracker,
respectively. The variable $f_\text{brem}$ is required to be smaller
than 20\%. In the range
$0.8<\abs{\eta}<2.5$~\cite{Khachatryan:2015hwa}, the resolution is
dominated by the incomplete containment of the $5{\times}5$ crystal
matrix caused by the larger amount of tracker material in this
region. Therefore, detailed performance comparisons are restricted to
events with electromagnetic showers occurring in the central region of
the EB.

\begin{figure}[!htbp]
\centering
\includegraphics[width=0.49\textwidth]{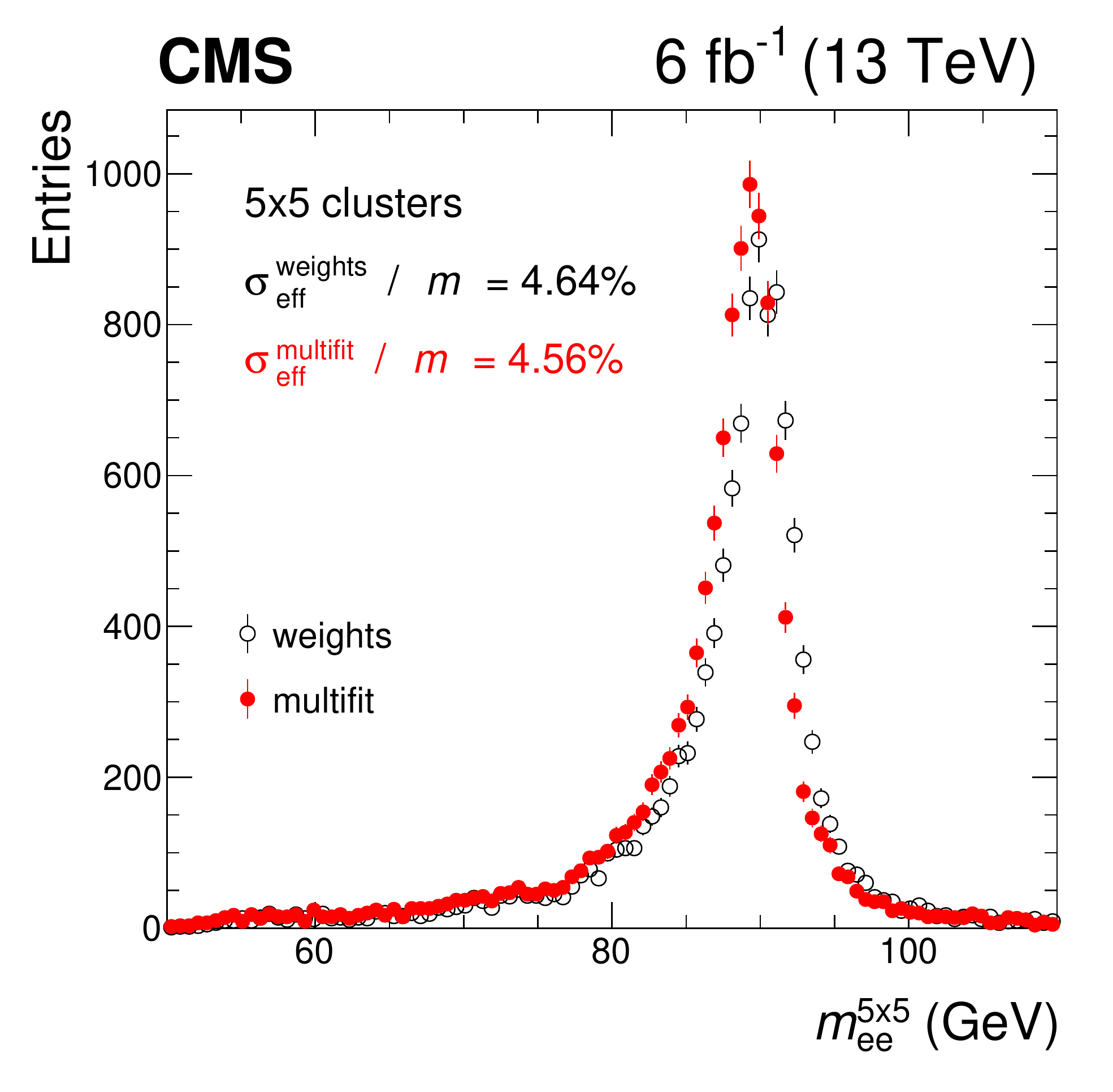}
  \caption{Example of the $\PZ\to\Pep\Pem$ invariant mass distribution
  in a central region of the barrel
  ($0.200<\text{max}(\abs{\eta_1},\abs{\eta_2})<0.435$) with the
  single-crystal amplitude estimated using either the weights or the
  multifit method.  A portion of collision data with typical Run~2
  conditions, recorded during October 2016, is used. Error bars
  represent the statistical uncertainty.  The energy is summed over a
  $5{\times}5$ crystal matrix. The reported values of
  $\sigma_\text{eff}$ include the natural width of the $\PZ$ boson,
  and are expressed as a percent of the position of the peak, $m$, of
  the corresponding invariant mass distribution.  \label{fig:resolZ}}
\end{figure}

Figure~\ref{fig:resolZ} shows the invariant mass of $5{\times}5$
cluster pairs, for a portion of the 2016 data, selecting pairs of
electrons, $\Pe_1$ and $\Pe_2$, that lie within a representative
central region of the barrel
($0.200<\text{max}(\abs{\eta_1},\abs{\eta_2})<0.435$). The outcome is
similar in other regions with low tracker material.  The shift in the
absolute energy scale for the simplified $5{\times}5$ clustering,
caused by the multifit $A_j$ being nonnegative for each BX, is not
corrected for.  The improvement is still significant for the $\pt$
range characteristic of $\PZ\to\Pep\Pem$ decays, matching the
expectation from the simulation, shown in Fig.~\ref{fig:resolution},
namely an improvement in resolution of $\approx$1\% in quadrature,
after unfolding the natural width of the $\PZ$ boson, for electrons
and photons with $30<\pt<100\GeV$.

A full comparison of the performance of the multifit algorithm in
Run~2 with that of the weights algorithm in Run~1 would require a
reanalysis of the Run~1 data, applying the more sophisticated
clustering techniques used in Run~2. Nevertheless, it is instructive
to make a straightforward comparison. For Run~1, where the crystal
energy was reconstructed with the default weights method, the electron
energy was estimated with the simple $5{\times}5$ crystal cluster, and
using the optimal calibrations of the 2012 data set ($\sqrt{s}$ =
8\TeV and 50\unit{ns} LHC bunch spacing)~\cite{Khachatryan:2015hwa}.
The effective resolution of the dielectron invariant mass
distribution, normalized to its peak, is
$\sigma_\text{eff}/m=4.59$\%. This is consistent with the value of
4.56\% obtained in Run~2 with the multifit algorithm, shown in
Fig.~\ref{fig:resolZ}. This indicates that the multifit method can
maintain the ECAL performance obtained during Run~1, in the \pt range
$\approx$(5--100)\GeV, relevant for most data analyses performed with
CMS, despite the substantially larger PU present in Run~2.

\subsubsection{Effect on jets}
\label{sec:jets}

The contribution to the average offset of the jet energy scale, from
the reconstructed electromagnetic component of each additional PU
interaction, was estimated in a simulated sample of pure noise in the
CMS detector by considering the energy contained in cones randomly
chosen within the detector acceptance. This shows that the
contribution to the offset from ECAL signals is reduced to a value of
less than 10\%, similar to that obtained in Run~1. Further details are
given in Ref.~\cite{CMS-DP-2018-028}.

\subsection{Reconstruction of cluster shape variables}
\label{sec:clustershape}

\begin{figure}[!htbp]
\centering
\includegraphics[width=0.49\textwidth]{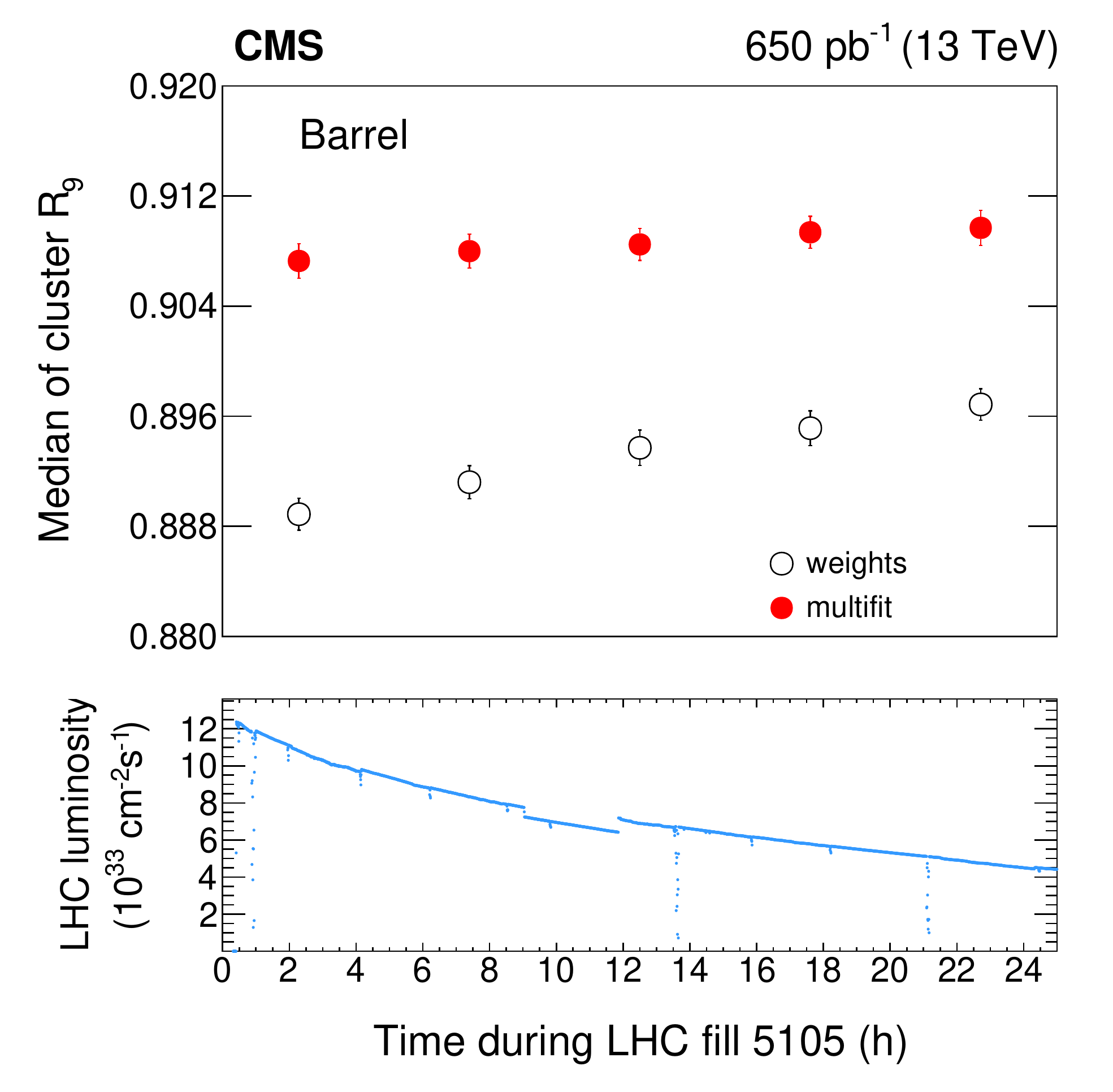}
 \caption{History of the median of the $R_9$ cluster shape for
 electrons from $\PZ\to\Pep\Pem$ decays during one typical LHC fill in
 2016. Hits are reconstructed with either the multifit (filled
 circles) or the weights algorithm (open circles). Each point
 represents the median of the distribution for a 5 hour period during
 the considered LHC fill. Error bars represent the statistical
 uncertainty on the median. The bottom panel shows the instantaneous
 luminosity delivered by the LHC as a function of time. The steps in
 the luminosity occurring about every two hours correspond to changes
 in the LHC beam crossing angle, which changes the overlap area of
 the bunches. Larger brief drops could indicate emittance scans during
 the fill. \label{fig:r9history}}

\end{figure}

The relative contribution of the PU energy within a cluster for
electrons from $\PZ$ boson decays is less than for clusters from
$\PGpz$ meson decays, and the sample of events is smaller. For these
reasons, it is difficult to estimate the variation of the energy scale
within one LHC fill arising from this contribution. The effect on the
cluster shapes is still significant, since they are computed using all
the hits in a cluster, including the low-energy ones. One example is
provided by the evolution, within an LHC fill, of the variable $R_9$,
defined as the ratio of the energy in a $3{\times}3$ crystal matrix
centered on the seed hit of the cluster, divided by the total energy
of the cluster.  This variable is an important measure of cluster
shape, since it is often used to distinguish between showering or
converted photons, and those not undergoing a bremsstrahlung process
or conversion within the tracker. For example, in studies of Higgs
boson physics, it is used to separate $\PH\to\PGg\PGg$ events into
categories with different $m_{\PGg\PGg}$ effective mass
resolutions.  Thus it is important that the $R_9$ variable remains
stable over time.  Figure~\ref{fig:r9history} shows the median of the
$R_9$ distribution for clusters from electron pairs in the barrel
having a mass consistent with that of the $\PZ$ boson, during an LHC
fill in 2016 with an average PU decreasing from a value of 42 at
the beginning of the fill to a value of 13 at the end.  The stability
of the cluster shape as a function of instantaneous luminosity,
obtained with the multifit algorithm, is clearly better than the one
obtained with the weights reconstruction.  The main reason the median
$R_9$ values drift up during a fill is that the denominator of the
$R_9$ ratio, which includes contributions from low-energy hits located
outside of the $3{\times}3$ matrix, decreases in the weights algorithm
when the instantaneous luminosity (and the PU) decreases.

Another effect that has been checked in data is the rejection power
for anomalous signals ascribed to direct energy deposition in the
APDs~\cite{Petyt:2012sf} by traversing particles. Unlike the hits in
an electromagnetic shower, the anomalous signals generally occur in
single channels of the calorimeter. They are rejected by a combination
of a topological selection and a requirement on the hit timing. The
topological selection rejects hits for which the value of the quantity
$(1-E_4/E_1)$ is close to 1, where $E_1$ is the energy of the crystal
and $E_4$ is the energy sum of the four nearest neighboring crystals.
A simulation of anomalous signals in the APDs is used, and the
efficiency is defined as the fraction of the reconstructed hits in
crystals with anomalous signals identified as such by the offline
reconstruction.  The rejection efficiency obtained when using the
multifit reconstruction is improved by as much as 15\% compared to the
weights method for hits with $E<15\GeV$. The probability of rejecting
hits from genuine energy deposits has been checked on data with hits
within clusters of $\PZ\to\Pep\Pem$ and is lower than $10^{-3}$ over
the entire $\pt$ spectrum of electrons from $\PZ$ boson decays for
both methods.

\section{Summary}
\label{sec:summary}

A multifit algorithm that uses a template fitting technique to
reconstruct the energy of single hits in the CMS electromagnetic
calorimeter has been presented. This algorithm was implemented before
the start of the Run~2 data taking period of the LHC, replacing the
weights method used in Run~1. The change was motivated by the
reduction of the LHC bunch spacing from 50 to 25\unit{ns}, and by the
higher instantaneous luminosity of Run~2, which led to a substantial
increase in both the in-time and out-of-time pileup.  Procedures have
been developed to provide regular updates of input parameters to
ensure the stability of energy reconstruction over time.

Studies based on $\PGpz\to\PGg\PGg$ and $\PZ\to\Pep\Pem$ control
samples in data show that the energy resolution for deposits ranging
from a few to several tens of \GeV is improved. The gain is more
significant for lower energy electromagnetic deposits, for which the
relative contribution of pileup is larger.  This enhances the
reconstruction of jets and missing transverse energy with the
particle-flow algorithm used in CMS. These results have been
reproduced with simulation studies, which show that an improvement
relative to the weights method is obtained at all energies, including
those relevant for photons from Higgs boson decays.

Simulation studies show that the new algorithm will perform
successfully at the high-luminosity LHC, where a peak pileup of about
200 interactions per bunch crossing, with 25\unit{ns} bunch spacing,
is expected.

\begin{acknowledgments}

  We congratulate our colleagues in the CERN accelerator departments for the excellent performance of the LHC and thank the technical and administrative staffs at CERN and at other CMS institutes for their contributions to the success of the CMS effort. In addition, we gratefully acknowledge the computing centers and personnel of the Worldwide LHC Computing Grid for delivering so effectively the computing infrastructure essential to our analyses. Finally, we acknowledge the enduring support for the construction and operation of the LHC and the CMS detector provided by the following funding agencies: BMBWF and FWF (Austria); FNRS and FWO (Belgium); CNPq, CAPES, FAPERJ, FAPERGS, and FAPESP (Brazil); MES (Bulgaria); CERN; CAS, MoST, and NSFC (China); COLCIENCIAS (Colombia); MSES and CSF (Croatia); RIF (Cyprus); SENESCYT (Ecuador); MoER, ERC IUT, PUT and ERDF (Estonia); Academy of Finland, MEC, and HIP (Finland); CEA and CNRS/IN2P3 (France); BMBF, DFG, and HGF (Germany); GSRT (Greece); NKFIA (Hungary); DAE and DST (India); IPM (Iran); SFI (Ireland); INFN (Italy); MSIP and NRF (Republic of Korea); MES (Latvia); LAS (Lithuania); MOE and UM (Malaysia); BUAP, CINVESTAV, CONACYT, LNS, SEP, and UASLP-FAI (Mexico); MOS (Montenegro); MBIE (New Zealand); PAEC (Pakistan); MSHE and NSC (Poland); FCT (Portugal); JINR (Dubna); MON, RosAtom, RAS, RFBR, and NRC KI (Russia); MESTD (Serbia); SEIDI, CPAN, PCTI, and FEDER (Spain); MOSTR (Sri Lanka); Swiss Funding Agencies (Switzerland); MST (Taipei); ThEPCenter, IPST, STAR, and NSTDA (Thailand); TUBITAK and TAEK (Turkey); NASU (Ukraine); STFC (United Kingdom); DOE and NSF (USA). 
  
  \hyphenation{Rachada-pisek} Individuals have received support from the Marie-Curie program and the European Research Council and Horizon 2020 Grant, contract Nos.\ 675440, 752730, and 765710 (European Union); the Leventis Foundation; the A.P.\ Sloan Foundation; the Alexander von Humboldt Foundation; the Belgian Federal Science Policy Office; the Fonds pour la Formation \`a la Recherche dans l'Industrie et dans l'Agriculture (FRIA-Belgium); the Agentschap voor Innovatie door Wetenschap en Technologie (IWT-Belgium); the F.R.S.-FNRS and FWO (Belgium) under the ``Excellence of Science -- EOS" -- be.h project n.\ 30820817; the Beijing Municipal Science \& Technology Commission, No. Z191100007219010; the Ministry of Education, Youth and Sports (MEYS) of the Czech Republic; the Deutsche Forschungsgemeinschaft (DFG) under Germany's Excellence Strategy -- EXC 2121 ``Quantum Universe" -- 390833306; the Lend\"ulet (``Momentum") Program and the J\'anos Bolyai Research Scholarship of the Hungarian Academy of Sciences, the New National Excellence Program \'UNKP, the NKFIA research grants 123842, 123959, 124845, 124850, 125105, 128713, 128786, and 129058 (Hungary); the Council of Science and Industrial Research, India; the HOMING PLUS program of the Foundation for Polish Science, cofinanced from European Union, Regional Development Fund, the Mobility Plus program of the Ministry of Science and Higher Education, the National Science Center (Poland), contracts Harmonia 2014/14/M/ST2/00428, Opus 2014/13/B/ST2/02543, 2014/15/B/ST2/03998, and 2015/19/B/ST2/02861, Sonata-bis 2012/07/E/ST2/01406; the National Priorities Research Program by Qatar National Research Fund; the Ministry of Science and Higher Education, project no. 02.a03.21.0005 (Russia); the Tomsk Polytechnic University Competitiveness Enhancement Program and ``Nauka" Project FSWW-2020-0008 (Russia); the Programa Estatal de Fomento de la Investigaci{\'o}n Cient{\'i}fica y T{\'e}cnica de Excelencia Mar\'{\i}a de Maeztu, grant MDM-2015-0509 and the Programa Severo Ochoa del Principado de Asturias; the Thalis and Aristeia programs cofinanced by EU-ESF and the Greek NSRF; the Rachadapisek Sompot Fund for Postdoctoral Fellowship, Chulalongkorn University and the Chulalongkorn Academic into Its 2nd Century Project Advancement Project (Thailand); the Kavli Foundation; the Nvidia Corporation; the SuperMicro Corporation; the Welch Foundation, contract C-1845; and the Weston Havens Foundation (USA). 
\end{acknowledgments}

\bibliography{auto_generated}

\cleardoublepage \appendix\section{The CMS Collaboration \label{app:collab}}\begin{sloppypar}\hyphenpenalty=5000\widowpenalty=500\clubpenalty=5000\vskip\cmsinstskip
\textbf{Yerevan Physics Institute, Yerevan, Armenia}\\*[0pt]
A.M.~Sirunyan$^{\textrm{\dag}}$, A.~Tumasyan
\vskip\cmsinstskip
\textbf{Institut f\"{u}r Hochenergiephysik, Wien, Austria}\\*[0pt]
W.~Adam, F.~Ambrogi, T.~Bergauer, M.~Dragicevic, J.~Er\"{o}, A.~Escalante~Del~Valle, R.~Fr\"{u}hwirth\cmsAuthorMark{1}, M.~Jeitler\cmsAuthorMark{1}, N.~Krammer, L.~Lechner, D.~Liko, T.~Madlener, I.~Mikulec, F.M.~Pitters, N.~Rad, J.~Schieck\cmsAuthorMark{1}, R.~Sch\"{o}fbeck, M.~Spanring, S.~Templ, W.~Waltenberger, C.-E.~Wulz\cmsAuthorMark{1}, M.~Zarucki
% ...
\vskip\cmsinstskip
\dag: Deceased\\
1:  Also at Vienna University of Technology, Vienna, Austria\\
% ...
\end{sloppypar}
\end{document}